\documentclass[12pt]{JHEP}

\font\blackboard=msbm10 at 12pt
\font\blackboards=msbm7
\font\blackboardss=msbm5
\newfam\black
\textfont\black=\blackboard
\scriptfont\black=\blackboards
\scriptscriptfont\black=\blackboardss
\def\bb#1{{\fam\black\relax#1}}

\newcommand{\NP}{{\em Nucl.\ Phys.\ }}

\newcommand{\PL}{{\em Phys.\ Lett.\ }}
\newcommand{\PR}{{\em Phys.\ Rev.\ }}

\newcommand{\PRL}{{\em Phys.\ Rev.\ Lett.\ }}

\newcommand{\ba}{\begin{array}}
\newcommand{\ea}{\end{array}}
\newcommand{\be}{\begin{equation}}
\newcommand{\ee}{\end{equation}}
\newcommand{\bea}{\begin{eqnarray}}
\newcommand{\eea}{\end{eqnarray}}
\newcommand{\beas}{\begin{eqnarray*}}
\newcommand{\eeas}{\end{eqnarray*}}

\def\beq{\begin{equation}}
\def\beqa{\begin{eqnarray}}
\def\eeq{\end{equation}}
\def\eeqa{\end{eqnarray}}

  \def\CD{{\cal D}} 
\def\e#1{{\rm e}^{^{\textstyle#1}}}

\def\darr#1{\raise1.5ex\hbox{$\leftrightarrow$}\mkern-16.5mu #1}

\def\half{{\textstyle{1\over2}}} 
\def\roughly#1{\raise.3ex\hbox{$#1$\kern-.75em\lower1ex\hbox{$\sim$}}}

\def\IB{\relax\hbox{$\inbar\kern-.3em{\rm B}$}}
\def\IC{\relax\hbox{$\inbar\kern-.3em{\rm C}$}}
\def\ID{\relax\hbox{$\inbar\kern-.3em{\rm D}$}}
\def\IE{\relax\hbox{$\inbar\kern-.3em{\rm E}$}}
\def\IF{\relax\hbox{$\inbar\kern-.3em{\rm F}$}}
\def\IG{\relax\hbox{$\inbar\kern-.3em{\rm G}$}}
\def\IGa{\relax\hbox{${\rm I}\kern-.18em\Gamma$}}
\def\IH{\relax{\rm I\kern-.18em H}}
\def\IK{\relax{\rm I\kern-.18em K}}
\def\IL{\relax{\rm I\kern-.18em L}}
\def\IP{\relax{\rm I\kern-.18em P}}
\def\IR{\relax{\rm I\kern-.18em R}}
\def\IZ{\relax\ifmmode\mathchoice{
\hbox{\cmss Z\kern-.4em Z}}{\hbox{\cmss Z\kern-.4em Z}}
{\lower.9pt\hbox{\cmsss Z\kern-.4em Z}}
{\lower1.2pt\hbox{\cmsss Z\kern-.4em Z}}
\else{\cmss Z\kern-.4em Z}\fi}
\def\II{\relax{\rm I\kern-.18em I}}

\def\ee#1{{\rm erf}\left(#1\right)}

\def\CD{{\cal D}}

\def\p{\partial}

\def\Tr{{\rm Tr}}

\def\inbar{\,\vrule height1.5ex width.4pt depth0pt}
\def\d{\delta}
\font\cmss=cmss10 \font\cmsss=cmss10 at 7pt

\def\d{{\delta}}

\def\e{{\epsilon}}

\def\m{{\mu}}
\def\n{{\nu}}

\def\s{{\sigma}}
\def\t{{\theta}}

\def\lref{\begingroup\obeylines\lr@f}
\def\lr@f#1#2{\gdef#1{\ref#1{#2}}\endgroup\unskip}
\def\s{\sigma}

\def\an{a_n}
\def\and{{a^\dagger_n}}

\def\P{\Psi}

\def\math@note#1{\gdef\@eqnlabel{LAB: #1}}
\title{Split string field theory II}
\author{David J. Gross\\
{Institute for Theoretical Physics}\\
{University of California, Santa Barbara}\\
{Santa Barbara, CA 93106, U.S.A.}\\
{\tt gross@itp.ucsb.edu}}
\author{Washington Taylor\footnote{Current
address:
Institute for Theoretical Physics,
University of California,
Santa Barbara, CA 93106-4030; {\tt  wati@itp.ucsb.edu}}\\
{Center for Theoretical Physics} \\
{MIT, Bldg.  6-306} \\
{Cambridge, MA 02139, U.S.A.} \\
{\tt wati@mit.edu}}

\abstract{We describe the ghost sector of cubic string field theory in
terms of degrees of freedom on the two halves of a split string.  In
particular, we represent a class of pure ghost BRST operators as
operators on the space of half-string functionals.  These BRST
operators were postulated by Rastelli, Sen, and Zwiebach to give a
description of cubic string field theory in the closed string vacuum
arising from condensation of a D25-brane in the original tachyonic
theory.  We find a class of solutions for the ghost equations of
motion using the pure ghost BRST operators.  We find a vanishing
action for these solutions, and discuss possible interpretations of
this result.  The form of the solutions we find in the pure ghost
theory suggests an analogous class of solutions in the original theory
on the D25-brane with BRST operator $Q_B$ coupling the matter and
ghost sectors.}

\preprint{MIT-CTP-3145, NSF-ITP-01-51, hep-th/0106036}
\begin{document}

\baselineskip16pt
\parskip=4pt
  \newpage

\section{Introduction}

From the   beginning  string theory was a theory of open strings.
It began as a model for an S-matrix \cite{Veneziano},
later interpreted as the S-matrix   of
open strings. Closed strings were discovered as singularities in the 
one-loop annulus
open string scattering diagram \cite{NGS}, and interpreted as new states in the
critical dimension (26 or 10) \cite{Lovelace}. This was  the first 
manifestation of one of
the new and most important features of string theory---namely what is 
now called
the UV-IR connection. Closed string states emerged not as dynamical 
bound states of
open strings, but rather appeared in closed string channels,
to lowest order in the string coupling, arising from  the ultraviolet 
behavior of  open
string propagators.

Once it became clear that string theory is a theory of quantum
gravity, and has many possible consistent classical backgrounds, it
became imperative to find a background-independent formulation of the
theory. A natural strategy was to construct a {\it string field
theory}, in which solutions of the classical string equations of
motion would correspond to different backgrounds, much as solutions of
Einstein's equations yield different background geometries in general
relativity. Witten's open string field theory raised hopes in this
direction, yielding a off-shell Chern-Simons-like theory of string
functionals \cite{Witten-SFT}, that reproduces open string scattering
amplitudes \cite{Giddings}.  This theory was generalized to open
superstrings \cite{Witten-super,Berkovits} and to a theory of open and
closed strings \cite{Zwiebach-open-closed}, although the formulation
of closed string field theory is much more complicated. The rather
complicated form of the explicit realization of string field theory,
however, made it difficult to treat the theory, even at the classical
level.

	Because open  string scattering amplitudes contain closed 
string poles, the
possibility has always existed that a complete formulation
of string theory might be given in terms of open
strings alone. This possibility has been greatly supported by recent 
developments.
First, the
web of dualities has revealed the unity of all previous formulations of string
theory. Closed string theories are connected, through such dualities, 
to open string
theories. Second, the discovery of D-branes has revealed the 
existence of new dynamical
objects
in closed string theory.
These D-branes can be thought of as solitons of open string theory, 
whose tension behaves
characteristically as $1/g^2_{\rm open}=1/g_{\rm closed}$, and whose
whose dynamics is controlled by open strings. Finally,   AdS/CFT 
duality provides a rich
class of examples whereby open string theory    captures the full 
essence of closed string
theory in a particular background \cite{Maldacena,gkp,Witten-AdS}.
The most remarkable feature of this duality is
that SUSY Yang Mills theory in four dimensions, the low-energy limit 
of the open string
theory on a D3-brane, is rich enough  to capture the physics of 
closed string theory in the
bulk of AdS$_5\times S^5$. The UV-IR connection plays an essential 
role in this scenario.

	The example of AdS/CFT duality once again raises the
possibility of giving a nonperturbative definition of string theory in
terms of open string theory, indeed, in this case, in terms of the low
energy limit of this theory---ordinary (supersymmetric) Yang-Mills
field theory. However, background independence seems out of reach in
this framework. To alter the spacetime background in the bulk requires
the insertion of operators of dimension greater than four in the gauge
theory. It is not known what regularization process is required or how
to handle such operators. Once again, to control the infra-red in the
closed string sector we need full control over the ultra-violet in the
open string sector. Furthermore, it is clear that to capture the full
moduli space of closed string backgrounds one would have to consider
extensions of gauge theory with an infinite number of
non-renormalizable interactions. Even a simple modification of the
background, such as turning on a background $B_{\m\n}$ field, requires
a major modification of the theory, namely a deformation that takes
one to a non-commutative field theory. Non-commutative field theory is
the one known example of a controllable deformation of gauge theory
that involves an infinite number of non-renormalizable interactions.
Standard field theoretic methods are  unable to deal with the
general case. String theory, however, knows how to control the
ultraviolet, so that the prospect remains that open string field
theory could capture the full essence of closed string theory,
including the full moduli space of closed string vacua.

  Open string field theory has recently come back into full focus, due
largely to the remarkable conjectures of Ashoke Sen, who suggested
that the tachyon of the ordinary 26-dimensional bosonic string theory
should be thought of as an instability of the space-filling D25-branes on
which this theory is constructed \cite{Sen-tachyon}. He argued that
the condensation of these tachyons defines a (classically) stable
vacuum of the theory in which the D-branes have annihilated, and that
about this vacuum configuration there should be no finite mass open
string excitations---this vacuum should instead describe the
26-dimensional closed string theory \cite{Sen-universality}.  Much
evidence in support of these conjectures has been given, using methods
of level truncation and using boundary  string field theory (see
paper I \cite{Gross-Taylor-I} for a list of references, and
\cite{Ohmori} for a review of this work). If true, these
conjectures strongly support the notion that open string theory is
powerful enough to contain classical D-brane solutions, and perhaps
the full content of string theory.

To realize these goals much remains to be done. First, one needs to 
achieve analytic
mastery of the classical string field theory equations of motion and
to construct analytic D-brane solutions. Second, one needs to show 
that at the stable
vacuum there are no open string physical excitations and that at the 
one-loop level the
ordinary closed string spectrum of physical states emerge.
Third, in order to achieve full background independence one needs to 
cast the formalism
of open string field theory into  an abstract framework, one that 
does not require
a particular closed string background, or a particular conformal 
field theory, for
its formulation. In such a  formalism the closed string background 
would emerge as a
consequence
of solving the equations of motion.

To this end, we attempt to give an operator formulation of string
field theory in which the string fields are defined as operators in an
appropriate space.  Roughly speaking the space is the space of
half-string states, and string fields, which are normally thought of
as functions of open strings $\Psi[x(\s)], \, 0\leq \s\leq \pi $ are
instead regarded as matrices $\Psi_{l,r}$, where $l(\s)=x(\s)$ and
$r(\s)=x(\pi -\s)$ for $ 0\leq \s\leq \pi/2$. With such an
identification Witten's $\star$ product is roughly described by matrix
multiplication.  Although such an interpretation was suggested in the
early papers on open string field theory
\cite{Witten-SFT,cubic,Horowitz-Strominger}, it appeared impossible to
precisely realize such a formalism because of difficulties associated
with the midpoint of the string. We are attempting to confront these
difficulties, starting with the very concrete problem of analytically
verifying Sen's conjectures, with the hope of then generalizing the
algebraic framework in a way that could be truly background
independent.  Other approaches to a half-string operator formalism for
open string field theory have appeared in
\cite{mo-tsun,bcnt,aab,Abdurrahman-Bordes,rsz-3,open-matrices}.

In our previous paper \cite{Gross-Taylor-I} (henceforth (I)), we
approached these issues in a simplified setting, following the
suggestion of Rastelli, Sen and Zwiebach \cite{rsz} that in the
locally stable vacuum in which the space-filling D25-brane has been
annihilated, it may be possible to describe the shifted string field
theory using a BRST operator $Q$ that is pure ghost.  If true, this
explains the decoupling of open strings at the stable vacuum; since
the cohomology of such a $Q$ is trivial, there are simply no
perturbative physical states. There can, of course, be physical
non-perturbative solitons, these should be the D-branes of the closed
string theory.

We would expect, even though there are no physical states in the
classical open string field theory around the closed string vacuum,
that closed string states should appear at one loop in off-shell open
string correlation functions. An interesting analogy to this situation
is QCD, in which there are no physical states corresponding to the
quark and gluon fields in the action, yet hadrons appear as physical
states in quark and gluon correlation functions. The situation here is
somewhat different. Quarks and gluons disappear from the physical
spectrum (are confined) only in a non-perturbative treatment of the
quantum theory---here the decoupling occurs at the classical
level. Hadrons only appear as non-perturbative bound states of quarks
and gluons---here closed strings should appear at one loop due to the
UV-IR connection. Another, perhaps useful analogy, is QCD$_2$, in
which, depending on the gauge choice or regularization of the gluon
propagator, quarks are either {\it i}. Infinitely massive and decouple
from the spectrum even off-shell (the {\it singular} gauge); or {\it
ii}. Have a finite mass but decouple on-shell from physical color
singlet hadrons (the {\it regular} gauge). The RSZ description of the
vacuum is analogous to the singular gauge where the decoupling of
quarks is trivial. There might very well be a less singular
description, in which off-shell open string states exist with finite
energy, but such that these states decouple on shell. In such a
description it might be much easier to find the closed string states,
since it is difficult (requires some kind of limiting procedure) to
obtain closed string states from the ultraviolet behavior of the open
string propagator when this propagator is singular.  Numerical
evidence \cite{Ellwood-Taylor,efhm} suggests that all open string
degrees of freedom decouple in the locally stable vacuum in Witten's
original description of the theory with BRST operator $Q_B$.  If we
can find an analytic description of this vacuum, this description of
the theory may turn out to be the best framework in which to
investigate the behavior of closed string degrees of freedom.

With a pure ghost $Q$, of the kind suggested by RSZ, the string field
equations of motion, $Q\Psi+\Psi \star \Psi=0$, can be factorized into
separate ghost and matter equations, when the full string field takes
the form $\Psi = \Psi_m \otimes \Psi_g$.  In the matter sector the
relevant equation is simply the projection equation,
$\Psi_m=\Psi_m\star\Psi_m$.  This factorization was used in
\cite{rsz-2} to identify certain solutions of the matter equation,
originally found in \cite{Kostelecky-Potting,Rastelli-Zwiebach}, as
the matter part of a D25-brane solution.  In (I) we developed a
split-string formalism for the matter sector of the theory, and showed
that these solutions of the matter equation are indeed projection
operators onto certain functionals on the space of half-string
configurations; related work appeared in \cite{rsz-3}.  A rank $N$
projection operator corresponds to a system of $N$ D-branes in the RSZ
model.  For example, a simple class of rank one projection operators
take the form $ e^{-l \cdot M \cdot l-r \cdot M \cdot r}$, where $l,
r$ describe the degrees of freedom of the left and right halves of the
string.  We showed that the sliver state discussed in
\cite{Kostelecky-Potting,Rastelli-Zwiebach,rsz-2} is a projection
operator of this form, and that for functionals independent of the
center of mass position of the string, rank $N$ projection operators
can be identified with solutions of the matter equation describing $N$
space-filling D25-branes.  Since all such projection operators are
equivalent under unitary transformations, all these solutions are
gauge-equivalent in the string field theory.  Projection operators
described by functionals which are localized in some subset of
space-time directions correspond to lower-dimensional D-branes.  We
showed that the sliver state describing a D-instanton can be modified
by a simple change in functional form so that it is only localized in
a subset of the space-time dimensions, and thus describes a
higher-dimensional D$p$-brane. In Section 2 we review the formulation
of Witten's cubic open string field theory and the results of (I). We
explain how the $\star$ product can be realized in the matter sector
as matrix (operator) multiplication and give the precise map between
this approach and the usual Fock space approach.  Some details of this
correspondence which did not appear in (I) are relegated to Appendices
\ref{sec:a-normalization}, \ref{sec:a-projector}.

  In Section 3 we turn to the split string formalism in the ghost
sector.  The situation here is trickier than in the matter sector for
several reasons.  First, the usual description of ghosts as fermionic
operators $c(\s), b(\s)$, does not lend itself easily to a natural
right-left split. The reason is that the $\star$ product does not
correspond to a simple overlap. In the matter sector the $\star$
product, $\Psi_1\star\Psi_2$, equates the right side of $\Psi_1$ with
the left side of $\Psi_2$ ($x_1(\s)=x_2(\pi-\s)$). However, for
functionals of $c(\s)$ the $\star$ product enforces the condition
$c_1(\s)=-c_2(\pi-\s)$.  Witten, in his original paper on the subject
\cite{Witten-SFT}, used a bosonized form of the ghosts. The bosonized
ghosts, $\phi(\s)$, do allow for the interpretation of the $\star$
product as identifying the left and right halves of the string,
however the BRST operator is then a non-local function of the bosonic
coordinates and hard to deal with. In \cite{Gross-Jevicki-1} the
bozonized ghosts were used to verify some of the properties of the
string action, however the full verification of the axioms
\cite{Gross-Jevicki-2} reverted to the use of fermionic ghosts, and
this is the formalism that has been primarily used in the intervening
years. We shall here use the bosonized form of the ghosts. In the
bosonized form the fermions correspond to kinks in the bosonic field,
and the operator $Q$ creates a kink at which the bosonic ghost field
$\phi$ has a jump discontinuity of magnitude $\pi$. This is a source
of many subtleties, but we think that these can be systematically
controlled.

	The other problem that arises in the ghost sector is the
string midpoint.  There is an anomaly due to the curvature of
the Riemann surface describing the three-string vertex. This leads to
an extra term in the $\star$ product that inserts ghost number, which
is equivalent to $\phi$ momentum, at the midpoint. The usual $Q$ acts
on the midpoint of the string, and this is a source of many
anomalies. In \cite{Gross-Jevicki-1,Gross-Jevicki-2} it was shown that
the anomalies all cancel in the critical dimension; but this subtlety
seems to be an obstacle to a simple realization of the $\star$ product
in an operator formalism. If, however, we use, following RSZ, a $Q$
formed solely out of ghosts we can construct $Q$ so that {\it it does
not act at the midpoint of the string}. This leads to an enormous
simplication and allows us to implement the operator formalism.
Remarkably, with this ansatz for the BRST operator, the problematic
anomalies are removed from the picture, even if we are not in the
critical dimension!

In Section 3 we establish the split field formalism in the ghost
sector and carefully show that a large class of $Q$'s, constructed
purely out of ghosts, satisfy all the axioms of string field theory,
when acting on well-behaved string fields, such as those corresponding
to states in the full string Hilbert space.

In Section 4 we briefly discuss the issue of how to describe the
physical, gauge invariant observables of open string field theory.
One of the main advantages of formulating string theory as a theory of
open strings is that it should be possible to give a precise list of
all possible gauge invariant observables without the problems that one
inevitably encounters in a closed string---quantum
gravitational---formulation. Thus, in the AdS/CFT duality the
observables of the boundary gauge theory are apparent; they are
correlation functions of local gauge invariant operators, that
correspond via the duality to S-matrix elements of the bulk
theory. But in addition, there are non-local gauge invariant
observables in the gauge theory, namely the expectation values of
Wilson loops, that correspond to some other kind of observables in the
bulk theory. Since, in principle, all local operators can be recovered
from Wilson loops, the complete set of physical observables in the
boundary theory are Wilson loops, which are themselves determined by
the gauge invariant eigenvalues of the covariant derivative that
effects parallel transport of gauge variant fields. In our operator
formulation of open string field theory we have a covariant derivative
operator, that transforms covariantly under gauge transformations, and
therefore we can define a large class of gauge invariant observables
(presumably a complete set of such observables) in terms of the
eigenvalues of this operator. This is discussed in Section 4.

In Section 5 we turn to the solution of the equations of motion in the
ghost sector.  We first discuss solutions in the ghost sector of the
projection equation $\Psi_g = \Psi_g \star \Psi_g$.  We then use the
ghost projector to construct a class of solutions to the full ghost
equation of motion $Q\Psi_g +\Psi_g \star \Psi_g = 0$ using pure ghost
operators $Q$ of the form described in Section 3.  In principle, these
solutions allow us to describe an arbitrary system of D$p$-branes in
the RSZ vacuum string field theory, when we combine the ghost
solutions with matter D$p$-brane solutions constructed in
\cite{rsz-3,Gross-Taylor-I}.  We find, however, that the action
apparently vanishes for the ghost solutions we construct.
Furthermore, due to the simplicity of the split string formalism, it
seems that for pure ghost operators $Q$ all solutions of the equations
of motion have vanishing action in this formalism.  We discuss
possible resolutions of this puzzle.  We also discuss a candidate
class of solutions to Witten's original cubic string field theory with
the BRST operator $Q_B$.  These solutions are related to a solution of
the form $Q_l | I \rangle$ proposed in \cite{cubic}, which takes the
theory to a purely cubic action, where, however, the identity state
$|I \rangle$ is replaced by a rank one projector.  It seems possible
that the solutions based on a rank one projector correspond to single
D-branes, while the solution based on the identity string field
corresponds to an infinite stack of space-filling D25-branes.  While
the action for all these solutions also formally vanishes in the split
string language, there are additional subtleties in this case arising
from the fact that the BRST operator has nontrivial action at the
string midpoint, so that the split string formalism is more
complicated than in the case of the simple ghost operators we describe
in Section 3.  We speculate that these additional complications may
lead to a finite value of the action for these solutions.


Finally, we present some conclusions and open problems    in Section 7.

\section{Review of String Field Theory}
\label{sec:review}

In this section we review some basic aspects of Witten's cubic string
field theory \cite{Witten-SFT}.  In subsection \ref{sec:SFT} we give a
general discussion of the theory using the formal language of
functional integrals, and in subsection \ref{sec:mode-matter} we
describe the matter sector of the theory in terms of Fourier modes on
the string.  In \ref{sec:splitting-matter} we describe the matter
sector of the theory in terms of split string degrees of freedom.
Appendices \ref{sec:a-normalization}, \ref{sec:a-projector}
describe in detail the connection between normalization factors in
the functional integral and Fock space descriptions of the theory.

\subsection{Witten's cubic string field theory}
\label{sec:SFT}

In \cite{Witten-SFT}, Witten proposed a simple formulation of  open
bosonic string field theory based on an action of Chern-Simons form
\begin{equation}
S = -\frac{1}{2}\int \Psi \star Q \Psi -\frac{g}{3}  \int \Psi \star
\Psi \star \Psi
\label{eq:SFT-action}
\end{equation}
where $g$ is the open string coupling and $\Psi$ is a
string field
taking values in a graded algebra ${\cal A}$.  Associated with the
algebra ${\cal A}$ there is a star product
\begin{equation}
\star:{\cal A} \otimes{\cal A} \rightarrow{\cal A}, \;\;\;\;\;
\end{equation}
under which the degree $G$ is additive ($G_{\Psi \star \Phi} = G_\Psi
+ G_\Phi$).  There is also a BRST operator
\begin{equation}
Q:{\cal A} \rightarrow{\cal A}, \;\;\;\;\;
\end{equation}
of degree one ($G_{Q \Psi} = 1 + G_\Psi$).  String fields can be
integrated using
\begin{equation}
\int:{\cal A} \rightarrow {\bb C}\,.
\end{equation}
This integral vanishes for all $\Psi$ with degree $G_\Psi \neq 3$.

The elements $Q, \star, \int$ defining the string field theory are
assumed to satisfy the following axioms:
\vspace*{0.2in}

\noindent {\bf (a)} Nilpotency of $Q$: $\;Q^2 \Psi = 0 \;\; \; \forall \Psi
\in{\cal A}$.

\noindent {\bf (b)} $\int Q\Psi = 0 \; \; \; \forall \Psi \in{\cal A}$.

\noindent {\bf (c)} Derivation property of $Q$:
$\;Q (\Psi \star \Phi) = (Q \Psi) \star \Phi +
(-1)^{G_\Psi} \Psi \star (Q \Phi) \; \; \; \forall \Psi, \Phi \in{\cal A}$.

\noindent {\bf (d)} Cyclicity:  $\;\int \Psi \star \Phi = (-1)^{G_\Psi
G_\Phi} \int \Phi \star \Psi \; \; \; \forall \Psi, \Phi \in{\cal A}$.
\vspace*{0.2in}

When these axioms are satisfied, the action (\ref{eq:SFT-action}) is
invariant under the gauge transformations
\begin{equation}
\delta \Psi = Q \Lambda + \Psi\star \Lambda - \Lambda \star \Psi
\end{equation}
for any gauge parameter $\Lambda \in{\cal A}$.

Witten presented this formal structure in \cite{Witten-SFT} and argued
that all the needed axioms are satisfied when ${\cal A}$ is taken to
be the space of string fields
\begin{equation}
{\cal A} =\{\Psi[x (\sigma); c (\sigma), b
(\sigma)]\}
\end{equation}
which can be described as functionals of the matter, ghost and
antighost fields describing an open string in 26 dimensions with $0 \leq \sigma \leq
\pi$.  For this string field theory, the BRST operator is the usual
open string BRST operator of the form
\begin{equation}
Q_B = \int_0^\pi d \sigma \; c (\sigma) \left(T^{(m)} (\sigma)
+\frac{1}{2}T^{(g)} (\sigma) \right)\,.
\end{equation}
The star product $\star$ is defined by gluing the right half of one
string to the left half of the other using a delta function
interaction.  The star product factorizes into separate matter and
ghost parts.  For the matter fields, the star product is given by
\begin{equation}
         \left(\P \star   \Phi\right) [z(\s)]
         \equiv
\int
\prod_{{0} \leq \s \leq {\pi\over 2}} dy(\s) \; dx (\pi -\sigma)
\prod_{{\pi\over 2} \leq
\s \leq \pi}
\delta[x(\s)-y(\pi-\s)]
\;   \P [x(\s)]  \Phi [y(\s)]
\label{eq:matter-star}
\end{equation}
\begin{eqnarray}
z(\s) & = &x(\s) \quad {\rm for} \quad {0} \leq \s \leq {\pi\over 2}\, ,
\label{eq:mult}\\
z(\s) & = &y(\s)\quad {\rm for} \quad   {\pi\over 2} \leq \s \leq \pi\, .
\nonumber
\end{eqnarray}
Similarly, the integral over a string field factorizes into matter and
ghost parts, and in the matter sector is given by
\begin{equation}
\int \Psi = \int \prod_{0 \leq \sigma \leq \pi} dx (\sigma) \;
\prod_{0 \leq
\s \leq \frac{\pi}{2} }
\delta[x(\s)-x(\pi-\s)] \;\Psi[x (\sigma)]\,.
\label{eq:integral-p}
\end{equation}

The ghost sector of the theory is defined in a similar fashion, but
has an anomaly due to the curvature of the Riemann surface describing
the three-string vertex.  The ghost sector can be described either in
terms of fermionic ghost fields $c (\sigma), b (\sigma)$ or through
bosonization in terms of a single bosonic scalar field $\phi
(\sigma)$.  From the functional point of view of (\ref{eq:mult},
\ref{eq:integral-p}), it is easiest to describe the ghost sector in the
bosonized language.  In this language, the star product in the ghost
sector is given by (\ref{eq:mult}) with an extra insertion of $\exp
(3i \phi (\pi/2)/2)$ inside the integral.  Similarly, the integration
of a string field in the ghost sector is given by (\ref{eq:integral-p})
with an insertion of $\exp (-3i \phi (\pi/2)/2)$ inside the integral.

The expressions (\ref{eq:mult}, \ref{eq:integral-p}) may seem rather
formal, however they can be given precise meaning when described in
terms of creation and annihilation operators acting on the string Fock
space.  This was done explicitly in
\cite{Gross-Jevicki-1,Gross-Jevicki-2,cst,Samuel,Ohta}, where the star
product and two- and three-string vertices were described in terms of
matter and ghost raising and lowering operators $a^\mu_n, b_n, c_n$.
In the following two subsections we describe the matter part of the
string field theory using a Fourier mode decomposition which replaces
the continuous set of degrees of freedom in the string $x (\sigma)$
with a set of modes $x_n$.  We describe the exact relationship between
this approach and the usual Fock space description of the theory in
Appendices \ref{sec:a-normalization} and \ref{sec:a-projector}.

\subsection{Mode decomposition of matter fields}
\label{sec:mode-matter}

We can expand each of the 26 matter fields $x^\mu$ in Fourier modes
through
\begin{equation}
x(\s) =  x_0+ \sqrt{2}\sum_{n=1}^\infty{x_n\cos(n\s)}\,.
\quad
\label{eq:x}
\end{equation}
(We will drop most spatial indices in this paper for clarity.)
The modes in (\ref{eq:x})  are related to creation and annihilation
operators through
\begin{eqnarray}
x_n  =  {i \over  {\sqrt{2n}} }
\left(a_n-a^\dagger_n\right) & \hspace{0.5in} &
            p_n =-i
{\p \over \p x_n} =  \sqrt{ n\over 2 } \left(a_n+a^\dagger_n\right)
         \label{eq:xp}\\
a_n = -i \left(  \sqrt{n\over 2} x_n + \frac{1}{ \sqrt{2n}}
\frac{\partial}{ \partial x_n}  \right) & &
a^{\dagger}_n = i \left(  \sqrt{ n\over 2} x_n  - \frac{1}{ \sqrt{2n}}
\frac{\partial}{ \partial x_n}  \right) \, ,\nonumber
\end{eqnarray}
for $n \neq 0$, and through
\begin{eqnarray}
x_0  =  {i \over  {{2}} }
\left(a_0-a^\dagger_0\right) & \hspace{0.5in} &
            p_0 =-i
{\p \over \p x_0} =   \left(a_0+a^\dagger_0\right)
         \label{eq:xp0}\\
a_0 = -i \left(   x_0 + \frac{1}{2}
\frac{\partial}{ \partial x_0}  \right) & &
a^{\dagger}_0 = i \left(   x_0  - \frac{1}{ 2}
\frac{\partial}{ \partial x_0}  \right)\, ,\nonumber
\end{eqnarray}
for the zero modes.
We write
\begin{eqnarray}
\quad |x)  & =&{i\over \sqrt{2}}E[|a)-|a^\dagger)] \\
         \quad |p) & = &
            {1\over \sqrt{2}E} [|a)+|a^\dagger)]\, , \nonumber
\end{eqnarray}
where
\begin{equation}
[\an,a^\dagger_m]=\delta_{nm}, \quad \quad    E^{-1}_{nm}=
\delta_{nm}\sqrt{n}+\delta_{n0}\delta_{m0} \sqrt{2} \, .
\label{eq:can}
\end{equation}

We can describe a string field (in the matter sector) as a functional
$\Psi[\{x_n\}]$ of the countable set of modes of the matter fields.
For such functionals, we can define the  string
field integral and star product so that
\begin{equation}
\int \Psi = \int \; \prod_{n = 0}^{\infty}
dx_n \;\prod_{k = 0}^{ \infty}  \delta (x_{2k + 1}) \;
\Psi[\{x_n\}]
\label{eq:integral-full}
\end{equation}
and
\begin{equation}
\int \Psi \star \Phi =
\int \; \prod_{n = 0}^{\infty}
dx_n \;
\Psi[\{x_n\}] \Phi[\{(-1)^nx_n\}]\,.
 \label{eq:integral-star}
\end{equation}
Throughout the paper we will use (\ref{eq:integral-full},
\ref{eq:integral-star}) to define the normalization of the string
field integral $\int$ and star product $\star$.

\subsection{Split string description of matter fields}
\label{sec:splitting-matter}

In this subsection we describe the splitting of the matter fields
into left and right components, as developed in
\cite{Gross-Jevicki-1,Gross-Taylor-I}.  Related approaches were
developed in \cite{mo-tsun,bcnt,Abdurrahman-Bordes,rsz-3}.

We split the string coordinate $x(\s)$ (which satisfies Neumann
boundary conditions at $\s =\ 0 , \pi$) into its left and right
pieces, according to
\begin{equation}
l(\s)=x(\s), \quad r (\s) = x (\pi -\s)
\;\;\;\;\; \quad{\rm for}\quad {{0} \leq \s \leq {\pi\over 2}} \ ,
\label{eq:defxrl}
\end{equation}
where $l(\s)$ and $r(\s)$ obey Neumann boundary
conditions at $\s =0$ and Dirichlet boundary conditions at $\s=\pi/2$.
We can perform a separate mode expansion on the left and right
pieces of the string
\begin{eqnarray}
            l(\s)&=&\sqrt{2}\sum_{n=0}^\infty
    l_{2n+1}
\cos(2n+1)\s \ ,
\label{eq:defrl2}\\
              r( \s)&= &\sqrt{2}\sum_{n=0}^\infty
    r_{2n+1}
\cos(2n+1)\s \ .\label{eq:r-expansion}
\end{eqnarray}
The half-string modes are related to the full-string modes through
\begin{eqnarray}
l_{2k+1}   &=
&x_{2k+1}+\sum_{n=0}^{\infty}X_{2k+1,2n}
\;x_{2n}  \, ,
\label{eq:xrel2}\\
r_{2k+1} & = &
-x_{2k+1}+\sum_{n=0}^{\infty}X_{2k+1,2n}
\;x_{2n}  \, .
\nonumber
\end{eqnarray}
and
\begin{eqnarray}
         x_{2n+1} &= &\half\left(l_{2n+1}-r_{2n+1}\right)  \, ,
\label{eq:xrel}\\
x_{2n}
   & = &  \half
\sum_{k=0}^{\infty}X_{2n,2k+1}\left(l_{2k+1}+r_{2k+1}\right)\, ,
\nonumber
\end{eqnarray}
where the transformation matrices $X_{2n, 2k + 1}, X_{2k + 1, 2n}$ are
given by
\begin{eqnarray}
X_{2k+1,2n}=X_{2n,2k+1} & = & { 4(-1)^{k+n}(2k+1)\over
\pi\left({(2k+1)^2-4n^2}\right)} \;\;\;  \quad  (n \neq 0)\, , \label{eq:X}\\
\quad X_{ 2k+1,0} =X_{0,2k+1} & = &
{  2 \sqrt{2}(-1)^{k}\over \pi{(2k+1)}}\, .\nonumber
\end{eqnarray}
The matrix
\begin{equation}
X \equiv  \pmatrix{  0 & X_{2k+1,2n} \cr
          X_{2n, 2k+1} & 0  \cr }  \,
\label{eq:matX}
\end{equation}
is symmetric and orthogonal: \ \  $X=X^T=X^{-1}$.

Using (\ref{eq:xrel}, \ref{eq:xrel2}) we can rewrite a string field
$\Psi[\{x_n\}]$ as a functional of the right- and left-half string
degrees of freedom $\Psi[\{l_{2k + 1}\};\{r_{2k + 1}\}]$.  Since $\det X
= 1$, we can then write the string field integral (\ref{eq:integral-full})
as
\begin{eqnarray}
\int \Psi & = & \int \; \prod_{k = 0}^{\infty}
\left(\frac{1}{2} \;dl_{2k + 1} \;dr_{2k + 1}
\,\delta ( \frac{l_{2k + 1}-r_{2k + 1}}{2} )\right)
\;\Psi[\{l_{2k + 1}\};\{r_{2k + 1}\}] \nonumber\\
  & = &
\int \; \prod_{k = 0}^{\infty}
  dl_{2k + 1}
\;\Psi[\{l_{2k + 1}\};\{l_{2k + 1}\}]\,. \label{eq:half-integral}
\end{eqnarray}
Similarly, the star product $\Psi \star \Phi$ is given in the split
string language by
\begin{equation}
(\Psi \star \Phi)[\{l_{2k + 1}\};\{r_{2k + 1}\}]=
\int  \; \prod_{k = 0}^{\infty}
  d y_{2j + 1}\;
\Psi[\{l_{2k + 1}\};\{y_{2j + 1}\}] \;
\Phi[\{y_{2j + 1}\};\{r_{2k + 1}\}]\,. \label{eq:half-star}
\end{equation}

This gives us a complete formulation of the matter part of the cubic
string field theory in terms of modes on the half-string.  In this
formulation, the string field $\Psi[l; r]$ essentially acts as an
operator on a space of half-string functionals
\begin{equation}
\Psi \Rightarrow \hat{\Psi}\,.
\end{equation}
In this operator language, the string field integral
(\ref{eq:half-integral}) and star product (\ref{eq:half-star}) can be
described in terms of a trace and operator multiplication respectively
\begin{eqnarray}
\int \Psi & \Rightarrow & {\rm Tr}\; \hat{\Psi} \\
\Psi \star \Phi & \Rightarrow &  \hat{\Psi} \hat{\Phi}\,.
\end{eqnarray}

One subtle aspect of this split string formalism involves the  role of
the string midpoint.  As we will discuss further  in Section
\ref{sec:ghosts}, these issues are more significant in the ghost
sector of the theory.  One way of dealing with the midpoint is to
explicitly subtract out the midpoint  before using (\ref{eq:xrel2}) to
relate the full-string modes to half-string modes.  We have then
\begin{eqnarray}
\tilde{x} (\sigma) & = &  x (\sigma) -\bar{x}\\
\tilde{x}_{n} & = &  x_{n} -\delta_{n0} \bar{x}
\end{eqnarray}
where
\begin{eqnarray}
\bar{x} & = & x (\pi/2) = x_0 + \sqrt{2}
\sum_{n = 1}^{ \infty}
(-1)^nx_{2n}
\end{eqnarray}
is the string midpoint.  Plugging $\tilde{x}_n$ into (\ref{eq:xrel2})
gives us a set of left and right modes $\tilde{l}_{2k + 1},
\tilde{r}_{2k + 1}$ associated with a string having vanishing
midpoint.  This gives us a representation of the string field $\Psi$
as an operator-valued function on space-time
\begin{equation}
\Psi \Rightarrow \hat{\Psi} = \tilde{\Psi}(\bar{x})
 \label{eq:operator-function}
\end{equation}
where at each point $\bar{x}$, the operator $\tilde{\Psi}(\bar{x})$
acts on the usual Hilbert space of states of a string with
Neumann-Dirichlet boundary conditions.  For a string field of definite
momentum $p_0$, the associated operator-valued function takes the form
\begin{equation}
\tilde{\Psi} (\bar{x}) = e^{ip_0 \bar{x}} \tilde{\Psi}_0
 \label{eq:p-operator}
\end{equation}
where $\tilde{\Psi}_0$ is a single operator acting on the ND Hilbert
space.  We will find this description of string fields with definite
momentum useful in the later discussion of the ghost sector of the
theory.

A class of matter string fields of particular interest are those which
satisfy the projection equation
\begin{equation}
\Psi = \Psi \star \Psi\,.
\label{eq:matter-projection}
\end{equation}
As discussed in \cite{Kostelecky-Potting}, such states could be a
first step towards an analytic construction of a solution to the full
string field equation of motion $Q\Psi + g \Psi \star \Psi = 0$.
Furthermore, in the vacuum string field theory postulated by Rastelli, Sen,
and Zwiebach, such states correspond to D-brane solitons
\cite{rsz,rsz-2}.  In \cite{Gross-Taylor-I,rsz-3} it was shown that
the projection operators associated with single D-branes are rank one
projectors on the appropriate space of string functionals when
described in the split string language; multiple D-brane
configurations are described by higher-rank projections.  In the
notation we are using here, a simple class of rank one projectors take
the form
\begin{equation}
\Psi[l; r] = \left(\det  \frac{M}{\pi}  \right)^{26/2}
\exp \left( \sum_{k = 0}^{ \infty}
-\frac{1}{2}l_{2k + 1} M_{2k + 1, 2j + 1} l_{2j + 1}
-\frac{1}{2}r_{2k + 1} M_{2k + 1, 2j + 1} r_{2j + 1} \right)\,.
\label{eq:Gaussian-projection}
\end{equation}
The matrix $M$ giving one such state associated with a D-instanton was
explicitly analyzed in \cite{Gross-Taylor-I} and related to the
``sliver state'' originally found in
\cite{Kostelecky-Potting,Rastelli-Zwiebach}.  From
(\ref{eq:half-star}) we see that the state
(\ref{eq:Gaussian-projection}) is indeed a projection operator
satisfying (\ref{eq:matter-projection}), which is
$\hat{\Psi}\hat{\Psi} = \hat{\Psi}$ in the operator language.  From
(\ref{eq:half-integral}) we see that furthermore ${\rm Tr}\;
\hat{\Psi} = 1$ for the state (\ref{eq:Gaussian-projection}), so that
this is actually a rank one projection operator.  In Appendices
\ref{sec:a-normalization} and \ref{sec:a-projector} we derive the
precise relations between the normalization factors in the functional
and Fock space approaches, and clarify the relationship between the
normalization we use here for projection operators and that used in
the discussion of \cite{rsz-3}.

\section{Split String Ghosts}
\label{sec:ghosts}

We now discuss the split string approach to the ghost sector of the
theory.  As discussed in the introduction, there are several
complications which make the ghost sector of the theory more difficult
to treat in this language than the matter sector.  In the fermionic
language, the overlap condition and anomalous midpoint insertions are
rather difficult to treat in a simple fashion.  These problems become
much simpler in the bosonized formulation of the ghost sector, in
which the ghosts are associated with a single additional bosonic field
with half-integral momentum.  In this section we give a detailed
discussion of the bosonized description of the ghost sector, and show
that the split string formalism of the matter sector carries over
immediately to the split string description of the string field star
product $\star$ and integral $\int$ for the bosonized ghosts.  The
remaining complication is the BRST operator $Q$.  In this paper we
concentrate on a particularly simple class of pure ghost BRST
operators which have no action on the string midpoint, and for which
the split string description is therefore particularly simple.  This
is the class of ghost operators which was proposed for the vacuum
string field theory in \cite{rsz}.  We show that these operators
satisfy all the axioms of string field theory when acting on suitably
well-behaved string states.  

In subsection \ref{sec:bosonization} we describe in detail the
bosonization of the fermionic ghost sector.  We give explicit formulae
relating the ghost fields in the two formalisms, and we discuss a
regulator that controls certain divergences which appear when doing
calculations with ghost fields.  In subsection
\ref{sec:splitting-ghosts} we discuss the star product and string
field integral in the bosonized ghost language.  These operations are
exactly the same as those we have discussed in the matter sector
except for the insertion of a midpoint-dependent phase factor.  In
\ref{sec:ghost-Q} we describe in detail the class of pure ghost BRST
operators of interest, and show that these operators indeed obey the
axioms of string field theory when acting on states in the original
string Fock space.

\subsection{Ghost bosonization}
\label{sec:bosonization}

In this section we review some basic aspects of the bosonization of
the ghosts in the open bosonic string.  Most
of this material is described in \cite{GSW}

\subsubsection{Fermionic ghosts}

The ghost and antighost fields on the open string satisfy periodic
boundary conditions
\begin{equation}
c^{\pm} (\sigma + 2 \pi) = c^{\pm} (\sigma), \;\;\;\;\;
b_{\pm} (\sigma + 2 \pi) = b_{\pm} (\sigma)\,.
\end{equation}
These Grassmann
fields have mode decompositions
\begin{eqnarray}
         c^{\pm}(\s)& = &
\sum_{n } c_ne^{\mp in\s}  \, \\
b_{\pm}(\s)& = & 
\sum_{n } b_ne^{\mp in\s}\, .
\label{eq:ghosts}
\end{eqnarray}
The ghost creation and annihilation operators satisfy
\begin{equation}
\{c_n ,b_m\} =\d_{n+m,0}\, ,\quad  \{c_n ,c_m\}=\{b_n ,b_m\}= 0 \, .
\label{eq:canghosts}
\end{equation}

The ghost Fock space has a pair of vacua $| \pm \rangle$ annihilated
by $c_n, b_n$ for $n > 0$.  These two vacua
satisfy
\begin{eqnarray*}
c_0 | -\rangle = | + \rangle & \hspace*{0.5in}  &
c_0 | + \rangle = 0  \\
b_0 | + \rangle = | -\rangle &  &  b_0 | -\rangle = 0\, .
\end{eqnarray*}
We shall  define the Fock space so that $b_0, c_0$ are hermitian and
$c^\dagger_n=c_{-n},\,
b^\dagger_n=b_{-n}$.
It follows that $\langle +|+\rangle=
\langle +|c_0|-\rangle =0$, and similarly $\langle -|-\rangle=0.$
We normalize the vacua so that
\begin{equation}
         \langle +|b_0|+\rangle= \langle -|c_0|-\rangle=1\, .
\end{equation}

The ghost number is
\begin{equation}
G=\sum_{n=1 }^\infty\left[ c_{-n}b_n-b_{-n}c_n\right]+\half \left[
c_{0}b_0-b_{0}c_0\right]  + 3/2
\end{equation}
so that the $c_n$ ($b_n$) have ghost number 1 (-1), for all $n$.  The
vacua $|  + \rangle$ and $| -\rangle$ have ghost number 2 and 1 respectively.

\subsubsection{Bosonized ghosts}

An alternative description of the fermionic ghost fields can be given
in terms of a single periodic bosonic field
\begin{equation}
\phi (\sigma) = \phi_0 + {\sqrt{2} }\sum_{n = 1}^{ \infty}
\phi_n \cos (n \sigma)\,.
\end{equation}
This field can be decomposed into right-moving and left-moving parts
$\phi^{\pm}$ and modes $a_n, a^{\dagger}_n$ through
\begin{eqnarray}
\phi^{\pm} (\sigma) & = &  \phi_0 \pm\sigma p_0
+\sum_{n = 1}^{ \infty}
\left[\sqrt{2}\phi_n \cos (n \sigma)
\pm\frac{\sqrt{2}p_n}{n}  \sin (n \sigma) \right]\\
    & = & \phi_0 \pm\sigma p_0
+i\sum_{n = 1}^{ \infty}
\left[-\frac{1}{\sqrt{n}}  e^{\pm in \sigma} a^{\dagger}_n
+\frac{1}{\sqrt{m}}  e^{\mp im \sigma} a_m\right]
\end{eqnarray}
where we use the conventions of Section 2 for the definitions of $a_n,
a^{\dagger}_n,$ and $p_n$.
\begin{equation}
[a_n, a_m^{\dagger}] =
\delta_{n, m},\quad \phi_n={i\over \sqrt{2n}}(a_n-a^{\dagger}_n), \quad p_n=
\sqrt{n\over 2}(a_n+a^{\dagger}_n)=-i{\p\over \p \phi_n}\, .
\end{equation}
We will often write $a_{-n} = a_n^{\dagger}$.

The fermionic ghosts are related to the bosonized ghost field
through the bosonization formulae \cite{Siegel-Zwiebach,GSW,Witten-SFT}
\begin{eqnarray}
c^+ (\sigma) & = & : e^{i \phi^+ (\sigma)}:
= e^{i \phi_0} e^{i \sigma (p_0 + 1/2)}
e^{\sum_{n = 1}^{ \infty} \frac{1}{\sqrt{n}} e^{in \sigma} a^{\dagger}_n }
e^{-\sum_{m = 1}^{ \infty} \frac{1}{ \sqrt{m}} e^{-im \sigma} a_m }
\label{eq:cp}\\ \nonumber \\
c^- (\sigma) & = & : e^{i \phi^- (\sigma)}:
= e^{i \phi_0} e^{-i \sigma (p_0 + 1/2)}
e^{\sum_{n = 1}^{ \infty} \frac{1}{\sqrt{n}} e^{-in \sigma} a^{\dagger}_n }
e^{-\sum_{m = 1}^{ \infty} \frac{1}{ \sqrt{m}} e^{im \sigma} a_m }
\label{eq:cm}\\
\nonumber\\
b_+ (\sigma) & = & : e^{-i \phi^+ (\sigma)}:
= e^{-i \phi_0} e^{-i \sigma (p_0 -1/2)}
e^{-\sum_{n = 1}^{ \infty} \frac{1}{\sqrt{n}} e^{in \sigma} a^{\dagger}_n }
e^{\sum_{m = 1}^{ \infty} \frac{1}{ \sqrt{m}} e^{-im \sigma} a_m }
\label{eq:bp}\\
\nonumber\\
b_- (\sigma) & = & : e^{-i \phi^- (\sigma)}:
= e^{-i \phi_0} e^{i \sigma (p_0 -1/2)}
e^{-\sum_{n = 1}^{ \infty} \frac{1}{\sqrt{n}} e^{-in \sigma} a^{\dagger}_n }
e^{\sum_{m = 1}^{ \infty} \frac{1}{ \sqrt{m}} e^{im \sigma} a_m }
\label{eq:bm}
\end{eqnarray}
The appearance of the extra factors of $e^{\pm i \sigma/2}$ in these
formulae indicates that for $c^{\pm}, b_{\pm}$ to be periodic we must
consider states described by functionals $\Psi[\phi (\sigma)]$ with
half-integral $p_0$, so that
\begin{equation}
\Psi[\phi (\sigma) + 2 \pi] = -\Psi[\phi (\sigma)]\, .
\end{equation}

{}From these formulae we can compute the anticommutators
\begin{eqnarray}
\{c^+ (\sigma), c^+ (\tau)\} & = &
0\label{eq:anticommutator1}\\
\{c^- (\sigma), c^- (\tau)\} & = & \{b_+ (\sigma), b_+ (\tau)\}
=\{b_- (\sigma), b_- (\tau)\} = 0 \label{eq:anticommutator2}\\
\{c^+ (\sigma),  b_+ (\tau)\} & = &
   2 \pi \delta (\sigma -\tau) \label{eq:anticommutator3} \, .
\end{eqnarray}

There are several approaches one can take to computing the
anticommutators (\ref{eq:anticommutator1}-\ref{eq:anticommutator3}).
We now describe one of these approaches in detail for the relation
(\ref{eq:anticommutator3}), illustrating a method of regulating
divergences which we will use throughout the paper.

In many computations involving operators like (\ref{eq:cp}) divergent
infinite sums arise.  One way of regulating these sums is to replace
the canonical commutation relations for the modes $[a_n, a_m^{\dagger}] =
\delta_{n, m}$ with the commutation relations
\begin{equation}
\left[ a_n,a_{m}^{\dagger}\right] = x^n \delta_{n, m}
\label{eq:regulator}
\end{equation}
where $x = 1-\epsilon < 1$.
With this regulator $[p_n,\phi_n] = -i x^n$, so the   regulated
form of $p_n$ is $p_n \to -ix^n\p/\p\phi_n$, which amounts to
replacing
$p_n
\rightarrow x^np_n$. Using this regulator,
from (\ref{eq:cp}, \ref{eq:bp}) we have
\begin{eqnarray}
\{c^+ (\sigma),  b_+ (\tau)\} & = &
   \left(
\frac{e^{\frac{i}{2}  (\sigma -\tau)}}{ 1-x e^{i (\sigma -\tau)}}
+\frac{e^{\frac{i}{2}  (\tau -\sigma)}}{ 1-x e^{i (\tau -\sigma)}}
\right) :c^+ (\sigma) b_+ (\tau): \label{eq:regulated-ac}
\end{eqnarray}
where
\begin{equation}
:c^+ (\sigma) b_+ (\tau): =
e^{ip_0 (\sigma -\tau)}
e^{\sum_{n = 1}^{\infty}\frac{1}{ \sqrt{n}}
(e^{in \sigma} -e^{in \tau})  a_n^{\dagger} }
e^{\sum_{m = 1}^{\infty}\frac{1}{ \sqrt{m}}
(e^{-im \tau} -e^{-im \sigma})  a_m }\,.
\label{eq:normal-cb}
\end{equation}
The quantity in parentheses in (\ref{eq:regulated-ac}) clearly
vanishes as $x \rightarrow 1$ when $\sigma \neq \tau$.  Integrating
this quantity, we see that as $x \rightarrow 1$, we have
\begin{equation}
\int_{\sigma =\tau -\pi}^{\sigma =\tau  + \pi}
\left(
\frac{e^{\frac{i}{2}  (\sigma -\tau)}}{ 1-x e^{i (\sigma -\tau)}}
+\frac{e^{\frac{i}{2}  (\tau -\sigma)}}{ 1-x e^{i (\tau -\sigma)}}
\right)
= \frac{-4i}{\sqrt{x}  }\ln
\left( \frac{1 + i \sqrt{ x}}{1-i  \sqrt{ x}}  \right)
   \rightarrow 2 \pi\, .
\end{equation}
Since (\ref{eq:normal-cb}) goes to 1 at $\sigma = \tau$, we see that
the anticommutator (\ref{eq:regulated-ac}) goes to $2 \pi \delta
(\sigma -\tau)$ in the limit $x \rightarrow 1$.  Note that this
argument involves some implicit assumptions about the nature of the
state the operator is acting on.  When (\ref{eq:normal-cb}) acts on a
string functional which is outside the Fock space and is sufficiently
poorly behaved, it is possible that the limit of this operator as $x
\rightarrow 1$ may not give 1.  Technically, relations
(\ref{eq:anticommutator1}-\ref{eq:anticommutator3}) are only valid
when these operators act on a well-behaved string functional such as
those in the Hilbert space associated with the bosonized ghost field.
We will return to this point in subsection
\ref{sec:ghost-Q}.

{}From (\ref{eq:cp}-\ref{eq:bm}) we can expand in modes to find an
expression for how any fermionic mode $c_n, b_m$ acts on a state of
fixed ghost number
in terms of the bosonic
modes $a_n$.  This expression depends upon the ghost number of the
state being acted on.  For example,
acting on states with $p_0 = -1/2$ we have
\begin{equation}
c_0 = e^{i \phi_0} \left[ 1 + a_{-1} a_1 +\frac{1}{2\sqrt{2}} a_{-1}^2
a_{2} +\frac{1}{2\sqrt{2}} a_{-2} a_{1}^2 +\frac{1}{2} a_{-2} a_{2} +
\cdots \right].
\label{eq:c01}
\end{equation}
Acting on states with $p_0 = 1/2$, on the other hand,
$c_0 = e^{i \phi_0} [ a_1 + \cdots ]$.
{}From these expressions, the action of the ghost and antighost modes on
any Fock space state can be computed.  Furthermore, the correspondence
between states in the bosonized and fermionic ghost Fock spaces can be
determined in this fashion.  For example, we have
\begin{eqnarray}
| -\rangle \;\;\; & \leftrightarrow & \;\;\; | p_0 = -1/2 \rangle
\nonumber\\
| + \rangle =c_0| -\rangle \;\;\; &
\leftrightarrow & \;\;\; | p_0 = 1/2 \rangle  \label{eq:states}\\
b_{-1}| -\rangle\;\;\; & \leftrightarrow &   \;\;\;| p_0 = -3/2 \rangle \, .
\nonumber
\end{eqnarray}

The bosonic and fermionic representations of the ghosts can also be
related using the formula
\begin{equation}
J_+ (\sigma) = \partial_\sigma \phi^+ (\sigma) =
\lim_{\tau \rightarrow \sigma}
\left[c^+(\sigma) b_+ (\tau)
+ \frac{i}{ \sigma -\tau}  \right]\, ,
\end{equation}
from which it follows that
\begin{eqnarray}
a_n &=& \frac{1}{\sqrt{|n |}}  \sum_{m}^{}  c_mb_{n-m}\, \, , n\neq 0
\nonumber \\
p_0&=&\sum_{n=1 }^\infty\left[ c_{-n}b_n-b_{-n}c_n\right]+\half \left[
c_{0}b_0-b_{0}c_0\right]  \ .
\end{eqnarray}
Note that $G = p_0 + 3/2$.

We will often find it convenient to express the ghost fields in terms
of the modes $\phi_n$ and derivatives $\partial/\partial \phi_n$.
Using the relations
    \begin{eqnarray}
\lefteqn{\exp\left[  \frac{1}{\sqrt{n}} e^{in \sigma} a^{\dagger}_n
\right]\exp\left[   -\frac{1}{\sqrt{n}}
e^{-in
\sigma} a_n \right]}\\
& = &\exp\left[
{1\over 2n}-i{ \sin(2n\s )\over
2n}\right]
\exp\left[   \sqrt{2}\,  \  {\sin(n\s)\over n} {\p \over \p \phi_n}\right]
\exp\left[   \sqrt{2}i\,{\cos(n\s) } \phi_n \right]
\\
& = &\exp\left[   {1\over 2n}+i{ \sin(2n\s )\over
2n}\right]
\exp\left[   \sqrt{2}i\,{\cos(n\s) } \phi_n \right]\exp\left[
\sqrt{2}\, \  {\sin(n\s)\over n} {\p
\over
\p \phi_n}\right] ,
\end{eqnarray}
$c^\pm(\s)$, for example, can be written as
\begin{eqnarray}
c^\pm (\s)  & = & K e^{i \e(\s){\pi\over 4}}
e^{\pm i \sigma (p_0-1) }e^{i \phi(\s)}
e^{\pm
\sum_{n = 1}^{ \infty}      \sqrt{2}\, \  {\sin(n\s)\over n} {\p
\over
\p \phi_n} }\label{eq:ccc1}\\
&= &K e^{-i \e(\s){\pi\over 4}}   e^{\pm i \sigma p_0 }
e^{\pm\sum_{n = 1}^{ \infty}      \sqrt{2}\, \  {\sin(n\s)\over n} {\p
\over
\p \phi_n} }e^{i \phi(\s)},
\label{eq:ccc}
\end{eqnarray}
where we have used
\begin{equation}
\sum_{n=1}^\infty{\sin(2n\s)\over 2n}={\pi\over 4}{\e}(\s ) -{\s\over
2}\quad {\rm for} \quad -\pi \leq
\s \leq \pi \, ,
\label{eq:sinsum}
\end{equation}
\noindent and $K$ is an (infinite) constant, $K=\exp[ \sum_n1/ 2n]$.
In (\ref{eq:sinsum}) we have kept the factor of ${\e}(\s )= \pm 1 $,
for $\s >0, \s  <0$, since we will use
this formula sometimes for negative values of $\s$.
Using the regulator (\ref{eq:regulator}), this constant becomes
\begin{equation}
K_x=\exp \left( \sum_{n = 1}^{\infty}\frac{x^{n}}{2n}  \right)
= \frac{1}{\sqrt{1-x}} \,.
\end{equation}
The regulated form of (\ref{eq:sinsum}) is
\begin{equation}
\sum_{n=1}^\infty{\sin(2n\s) x^n\over 2n}= -{\s\over
2}+ \frac{1}{4i}  \ln \left[
\frac{e^{i \sigma} -e^{-i \sigma} + \epsilon \; e^{-i \sigma}}{
e^{-i \sigma} -e^{i \sigma} + \epsilon \; e^{i \sigma}}  \right]\,.
\end{equation}
This expression gives (\ref{eq:sinsum}) when $\epsilon\ll \sigma$.
For small $\epsilon \sim \sigma$, we can replace this with
\begin{equation}
\sum_{n=1}^\infty{\sin(2n\s)x^n\over 2n}\sim-{\s\over
2}+ \frac{1}{4i}  \ln \left[
\frac{\epsilon + 2i \sigma}{ \epsilon -2i \sigma}  \right]
\end{equation}
for a simpler but equivalent regulator.
The regulated form of (\ref{eq:ccc1}, \ref{eq:ccc}) is then
\begin{eqnarray}
c^\pm (\s)  & = & K_x 
\left( \frac{\epsilon + 2i \sigma}{ \epsilon -2i \sigma}  \right)^{1/4}
e^{\pm i \sigma (p_0-1) }e^{i \phi(\s)}
e^{\pm
\sum_{n = 1}^{ \infty}      \sqrt{2}\, \  {\sin(n\s)\over n} x^n{\p
\over
\p \phi_n} }\label{eq:cx1}\\
&= &K_x 
\left( \frac{\epsilon - 2i \sigma}{ \epsilon +2i \sigma}  \right)^{1/4}
e^{\pm i \sigma p_0 }
e^{\pm\sum_{n = 1}^{ \infty}      \sqrt{2}\, \  {\sin(n\s)\over n}x^n {\p
\over
\p \phi_n} }e^{i \phi(\s)},
\label{eq:cx2}
\end{eqnarray}

Using (\ref{eq:ccc}) we can evaluate the action of
$c^+(\s_0)$ on a position basis state $|\phi(\s)\rangle = | \{\phi_n\}
\rangle$, which is given by (here we set $x=1$)
\begin{eqnarray}
c^+ (\s_0)|\phi(\s)\rangle &=&K e^{-i {\e}(\s_0 ){\pi\over 4}}  e^{i
\phi(\s_0)}
e^{i \sigma_0 p_0 }
    |\phi(\s)-(\pi-\s_0)\theta(\s_0-\s)
+\s_0\theta(\s -\s_0)\rangle \nonumber \\
&= &K e^{-i {\e}(\s_0 ){\pi\over 4}}  e^{i \phi(\s_0)} |\phi(\s)- \pi
\theta(\s_0-\s)\rangle
\, ,
\label{eq:conphi}
\end{eqnarray}
where we have used:
\begin{equation}
2\sum_{n=1}^\infty {\sin(n\s_0)\cos(n\s)\over n}=
(\pi-\s_0)\theta(\s_0-\s)-\s_0\theta(\s -\s_0), \;\;\;\;\; 0 \leq
\sigma_0, \sigma \leq \pi
\, .
\label{eq:sincossum}
\end{equation}
Note that in this equation, we must interpret $\theta(0)$ as equal to
${1\over 2}$,
since when $\s_0=\s$,   (\ref{eq:sincossum}) is identical to twice
(\ref{eq:sinsum})
Similarly to (\ref{eq:conphi}), we have
\begin{eqnarray}
c^- (\s_0)|\phi(\s)\rangle &=&
K e^{-i{\e}(\s_0 ) {\pi\over 4}}  e^{i \phi(\s_0)} |\phi(\s)+ \pi
\theta(\s_0-\s)\rangle
\, .
\label{eq:conphi2}
\end{eqnarray}
Thus, we see that the operators $c^\pm (\sigma)$ introduce a jump
discontinuity of magnitude $\pi$ in a smooth string configuration.
The presence of such kinks is the key mechanism underlying the
encoding of the full structure of the fermionic ghost-antighost Fock
space in the bosonized ghost Fock space.

We can describe the action of $c^{\pm} (\sigma_0)$ on a general
functional $\Psi[\phi (\sigma)]$ through
\begin{equation}
c^{\pm} (\sigma_0)  \Psi [\phi (\sigma)]
= K e^{ i{\e}(\s_0 ) {\pi\over 4} }
e^{i \phi (\sigma_0)} \Psi[\phi (\sigma) \pm \pi \theta (\sigma_0-\sigma)]\, ,
\label{eq:c-psi}
\end{equation}
or,
\begin{equation}
   c^{\pm} (\sigma_0)  \Psi [\phi_0, \phi_n ]
= Ke^{ i{\e}(\s_0 ) {\pi\over 4} }
e^{i \phi (\sigma_0)} \Psi[\phi_0 \pm \s_0 , \phi_n\pm
\sqrt{2}{\sin(n\s_0)\over n } ]\, .
\label{eq:c-psicomp}
\end{equation}

If we wish to use the regulated form of $c^\pm$, we need the sum ($\e=1-x$)
\begin{eqnarray}
2\sum_{n=1}^\infty {\sin(n\s_0)\cos(n\s)x^n\over n}
& = &-\s_0  +{1\over 2i}\ln\left[
\frac{(e^{i (\sigma_0 -\sigma)/2}-x e^{i (\sigma -\sigma_0)/2})
(e^{i (\sigma_0 +\sigma)/2}-x e^{-i (\sigma +\sigma_0)/2})
}{(e^{i (\sigma -\sigma_0)/2}-x e^{i (\sigma_0 -\sigma)/2})
(e^{-i (\sigma_0 +\sigma)/2}-x e^{i (\sigma +\sigma_0)/2})}
\right]
\,
\nonumber\\
  & \sim &
\pi/2-\s_0  +{1\over 2i}\ln\left[
\frac{\epsilon + i (\sigma_0-\sigma)
}{\epsilon -i (\sigma_0-\sigma)}
\right]\, ,
\label{eq:sincossumreg}
\end{eqnarray}
where in the second line we have simplified using $ \epsilon \sim
\sigma_0-\sigma\ll \sigma_0$.  This expression reduces to
(\ref{eq:sincossum}) when $\e \to 0$.  The regulated form of
(\ref{eq:c-psi}) is
\begin{equation}
c^{\pm} (\sigma_0)  \Psi [\phi (\sigma)]
= K_x 
\left( \frac{\epsilon + 2i \sigma}{ \epsilon -2i \sigma}  \right)^{1/4}
e^{i \phi (\sigma_0)} \;\Psi\left[\phi (\sigma) \pm  \pi/2\pm  {1\over
2i}\ln\left({\e +i(\s_0-\s)\over \e
-i(\s_0-\s)}\right) \right]\, .
\label{eq:c-psi-r}
\end{equation}

It might appear disturbing that the action of $c^\pm(\s)$ on a
functional is proportional to the infinite factor $K$. However this
factor is precisely what is needed to render the action of $c^\pm(\s)$
finite, when acting on a functional that lies within the ghost Fock
space.  Namely, when $c^\pm(\s)$ is smeared with a smooth function,
the action of the resulting operator on a Fock space functional gives
another Fock space functional. This may seem surprising, since $c^\pm(\s)$
creates a (fermionic) kink at $\s$; but when smeared the kinks average
out and the factor of $K$ cancels a corresponding vanishing factor of
$1/K$.

As an example of this phenomenon it is instructive to consider the
relation $c_0 | -\rangle = | + \rangle$.  From (\ref{eq:states}), we
see that in the bosonized language this becomes
\begin{equation}
{1\over 2 \pi}\int_0^\pi d \sigma \left(c^+ (\sigma) + c^-(\sigma)\right)
|{p_0 = -1/2} \rangle
=|{p_0 = 1/2} \rangle\,.
\end{equation}
This relation is straightforward to derive using the expressions for
$c^\pm$ in terms of raising and lowering operators (\ref{eq:cp},
\ref{eq:cm}).  It is somewhat less transparent, however, when we use
(\ref{eq:c-psi}) to describe the action of the $c$'s on the vacuum
wavefunction
\begin{equation}
\Psi_0[\phi (\sigma)] =
C \exp (-i \phi_0/2) \exp \left( -  \sum_{n = 1}^{ \infty}
   \frac{n}{2} \phi_n^2 \right)
\end{equation}
where
\begin{equation}
C =\prod_{n}\left( \frac{n}{\pi}  \right)^{26/4}\,.
\end{equation}
We have
\begin{eqnarray}
\lefteqn{c_0 \Psi_0[\phi (\sigma)]} \nonumber\\
& = &
\frac{1}{2 \pi} \int_0^\pi d \sigma
K C e^{i \e(\s)\pi/4} \exp \left( i \phi_0/2 + i \sum_{n}\sqrt{2}
\phi_n \cos (n
\sigma) \right)\nonumber\\
& &  \hspace*{0 in}\times
\left[
\exp \left( -i\sigma/2 -\sum_{n}\frac{n}{2}
(\phi_n + \frac{\sqrt{2}}{n} \sin n \sigma )^2 \right) +
\exp \left( i\sigma/2-\sum_{n}\frac{n}{2}
(\phi_n -\frac{\sqrt{2}}{n} \sin n \sigma )^2 \right) \right]
\nonumber
\\
& = &
C\exp \left(i \phi_0/2- \sum_{n}  \frac{n}{2}  \phi_n^2 \right)
  \int_{-\pi}^\pi \frac{d \sigma}{2 \pi}
\exp \left( -i \sigma/2 +i \e(\s)\pi/4 +  \sum_{n}
( i \sqrt{2} \phi_n e^{in \sigma}
+ \frac{\cos 2n \sigma}{ 2n}  )\right)   \nonumber\\
& = &
C\exp \left( i \phi_0/2- \sum_{n}  \frac{n}{2}  \phi_n^2 \right)
  \int_{-\pi}^\pi \; 
 \frac{d \sigma}{2 \pi}
\frac{\exp \left( 
-i \sigma/2 +i \e(\s)\pi/4  + \hskip -.02in\sum_{n}
   i \sqrt{2} \phi_n e^{in \sigma}
\right)}{  (2-2 \cos 2 \sigma)^{1/4}}
  \nonumber\\
& =  &C\exp \left(i \phi_0/2- \sum_{n}  \frac{n}{2}
\phi_n^2 \right) \, .
\label{eq:inth}
\end{eqnarray}
Here we used the result, derived in Appendix A.3, that the integral in
the next-to-last line equals one.

Thus we see that $c_0$, although it is an integral of operators that
create kinks, acts on the vacuum functional, which is a Gaussian about
$\phi_n=0$, to produce another Gaussian localized at $\phi_n=0$. The
kinks average out. The same will hold for any state in the Fock space.
Such states are of the form $P(\phi_n)\Psi_0[\phi (\sigma)]$, where
$P(\phi_n)$ is a polynomial in the $ {\phi_n}$'s.

\subsection{Splitting bosonic ghosts}
\label{sec:splitting-ghosts}

Using the bosonized formulation of the ghosts, we can use the same
approach to splitting the bosonic ghost field as was described in
Section \ref{sec:splitting-matter} for the matter fields; related work
appeared in \cite{aab}.  We are
particularly interested in ghost functionals $\Psi[\phi (\sigma)]$
with fixed ghost number $G = p_0 + 3/2$.  By separating out the string
midpoint $\bar{\phi}$ explicitly, as in (\ref{eq:operator-function},
\ref{eq:p-operator}), such functionals can be written as
operator-valued functions on space-time
\begin{equation}
\Psi[\phi (\sigma)] \Rightarrow \hat \P = 
e^{ip_0 \bar{\phi}} \tilde{\Psi}
\label{eq:map-phi}
\end{equation}
where $\tilde{\Psi}$ acts on the space of functionals
$\chi[\tilde{l}]$ of half-string bosonized ghost field configurations
$l (\sigma) = \phi (\sigma) -\bar{\phi}, 0 \leq \sigma \leq \pi /2$
vanishing at the midpoint.  In some situations we will separate out
the midpoint dependence of a state explicitly; in other situations it
is more convenient to keep the functional dependence on the midpoint
encoded in the left and right degrees of freedom through $\Psi[l; r]
\Rightarrow \hat{\Psi}$.

The star product in the ghost sector has an extra insertion of $\exp
(3i \bar{\phi}/2)$ due to the ghost current anomaly.  Thus, a pair of
states $A, B$ associated with operators $\hat{A}, \hat{B}$ through
(\ref{eq:map-phi}) have a star product given by
\begin{equation}
A \star B \Rightarrow \hat{A} e^{ \frac{3 i}{2} \bar{\phi}} \hat{B}
= e^{i \bar{\phi} (3/2 + p_{0 (A)}+p_{0 (B)})} \tilde{A} \tilde{B}\,,
\label{eq:bosonic-star}
\end{equation}
where $p_{0 (A, B)}+ 3/2$ are the ghost numbers of states $A, B$
respectively.  We will find it useful to write the midpoint factor
$e^{3i \bar{\phi}/2}$ appearing in this expression as
\begin{equation}
M =\exp \left( \frac{3i \bar{\phi}}{2} \right)\,.
\end{equation}
Since all ghost states of interest have definite ghost number $G$, and
therefore definite ghost momentum $p_0$ all the midpoint dependence of
the corresponding functionals are encoded in (\ref{eq:map-phi}).  It
follows that $M$ simply adds momentum to a state and therefore
commutes with all operators $\hat{A}$ associated with states of
definite ghost number.  In terms of $M$, we write
\begin{equation}
A \star B \Rightarrow \hat{A} M \hat{B}\,.
\end{equation}

The integral over a ghost string field $\Psi[\phi (\sigma)]$ is given
in this notation by
\begin{equation}
\int \Psi[\phi (\sigma)] \Rightarrow
{\rm Tr}\; M^{-1} \hat{\Psi} \equiv
\frac{1}{2 \pi} \int_0^{2 \pi} d \bar{\phi} \;\;{\rm Tr}_{\rm ND}\;
\left( M^{-1} \hat{\Psi} \right)\,
\end{equation}
where the trace ${\rm Tr}_{{\rm N D}}$ is taken in the ND half-string
space spanned by the basis $| \tilde{l} \rangle$\,.

\subsection{Ghostly $Q$'s}
\label{sec:ghost-Q}

We are now interested in constructing a class of operators $Q$ which
satisfy the axioms listed in Section \ref{sec:SFT} for the cubic
string field theory.  Following Rastelli, Sen and Zwiebach \cite{rsz},
we will consider operators which are linear in $c^{\pm}$.  We can
construct a general such operator of the form
\begin{equation}
Q_f = \int_0^\pi d \sigma \; \left(f_+ (\sigma) c^+ (\sigma)
+ f_-(\sigma) c^-(\sigma) \right)\, .
\label{eq:ghost-BRST}
\end{equation}

In order for $Q_f$ to define a consistent cubic string field theory,
we must show that
\vspace*{0.2in}

\noindent {\bf (a)} $Q_f^2 = 0$

\noindent {\bf (b)} $\int Q_f \Psi = 0 \; \;\forall \Psi$

\noindent {\bf (c)} $Q_f (A \star B) = (Q_f A) \star B + (-1)^{G_A}  A
\star (Q_f B)$,
where $G_A$ is the ghost number of  $A$.
\vspace*{0.2in}

We will restrict $f_\pm (\sigma)$ to satisfy the conditions
\begin{eqnarray}
f_\pm (\sigma) & = &  f_\pm (\pi -\sigma) \label{eq:restriction1}\\
f_\pm ( \pi/2) & = &  0\,.\label{eq:restriction2}
\end{eqnarray}
The first restriction, $f_\pm (\sigma) = f_\pm (\pi -\sigma)$, will be
necessary if $Q_f$ is to be a derivation and satisfy $\int Q_f\Psi=0$.
We require the second condition $f_\pm ( \pi/2) = 0 $ so as to avoid
problems at the midpoint.  Thus, for example, we do not consider
$f=1$, which would produce $Q_1= c_0$, but we do consider
$f=1+\cos(2\s)$, which produces $Q_{1+\cos(2\s)}= c_0+(c_2+c_{-2})/2$.
In \cite{rsz}, Rastelli, Sen and Zwiebach considered two classes of
pure ghost BRST operators.  The first class, including operators such
as $c_0$, do not annihilate the ghost identity field $ | I_g \rangle$,
and therefore do not satisfy the condition $\int Q \Psi = 0$.  The
second class of operators, including operators such as $c_0 + (c_2 +
c_{-2})/2$ do annihilate the identity and therefore satisfy the string
field theory axiom $\int Q \Psi = 0$.  As we will see, the restriction
(\ref{eq:restriction2}) amounts to the selection of BRST operators in
the second class.

We will now proceed to demonstrate each of the conditions (a-c) for
BRST operators given by (\ref{eq:ghost-BRST}) under the restrictions
(\ref{eq:restriction1}, \ref{eq:restriction2}), using the split string
formalism described in the last two subsections.  We shall not use the
regulator in this discussion for clarity, but for a situation in which
complications might be imagined we show explicitly in Appendix A.4
that the result still follows when the regulator is included, at least
when dealing with well-behaved functionals such as those in the string
Hilbert space.  We will work with a simplified BRST operator with
$f_-(\sigma) = 0$; the proofs for a general operator with $f_-\neq 0$
follow in a similar fashion.

We begin by describing the action of $Q_f$ on a state $\Psi[\phi
(\sigma)]$ with momentum $p_0$ explicitly in both the full- and
half-string formalisms. 
  From (\ref{eq:c-psi}), we have
\begin{equation}
Q_f \Psi[\phi (\sigma)]
= \int_0^\pi d \tau \;f_+ (\tau)
K  e^{i \frac{\pi}{ 4} }
e^{i \phi (\tau)} \Psi[\phi (\sigma) +  \pi \theta (\tau -\sigma)]\,.
\label{eq:regulated-q}
\end{equation}
In terms of the half-string variables, this becomes
\begin{eqnarray}
Q_f \Psi[l (\sigma), r (\sigma)] & = &
\int_0^{\pi/2} d \tau \;f_+ (\tau)
K  e^{i \frac{\pi}{ 4}  }
\left[
e^{i l (\tau)} \Psi[l (\sigma) +   \pi \theta (\tau -\sigma), r
(\sigma)]
\right.\\
& &\hspace{2in}
\left.
+e^{i r (\tau)} \Psi[l (\sigma) +  \pi, r (\sigma)
   +  \pi \theta (\sigma -\tau)] \right] \nonumber
\\
   & = &
\int_0^{\pi/2} d \tau \;f_+ (\tau)
K  e^{i \frac{\pi}{ 4}  }
\left[
e^{i l (\tau)} \Psi[l (\sigma) +  \pi \theta (\tau -\sigma), r
(\sigma)]
\right.\label{eq:split-action}\\
& &\hspace{2in}
\left.
+e^{ip_0 \pi} e^{i r (\tau)} \Psi[l (\sigma), r (\sigma)
   -  \pi \theta (\tau -\sigma)]\right] \nonumber
\end{eqnarray}
There is no difficulty in separating the action of $Q_f$ into left and
right parts because of the condition $f_+ (\pi/2) = 0$.
We can rewrite (\ref{eq:split-action}) as
\begin{equation}
Q_f \Psi \Rightarrow \hat{Q}_f \hat{\Psi} +
e^{i \pi (p_0 + 1/2)  } \hat{\Psi} \hat{Q}_f =\hat{Q}_f \hat{\Psi}
-(-1)^{G_\Psi} \hat{\Psi} \hat{Q}_f
\label{eq:qfsss}
\end{equation}
where
\begin{equation}
\hat{Q}_f = \int_0^{\pi/2} d \tau \;
f_+ (\tau) K  e^{i \frac{\pi}{ 4}  } e^{i\hat l (\tau)}
e^{i \pi \int_0^\tau d \rho \; \hat{p} (\rho)}\,.
 \label{eq:qh}
\end{equation}
The operator $\hat{Q}_f$ commutes with the midpoint insertion $e^{3i
\bar{\phi}/2} $,
\begin{equation}
 \hat{Q}_fM = M \hat{Q}_f\,.
\label{eq:qm-commute}
\end{equation}
Again, this follows immediately from the fact that we have restricted
attention to operators with $f_+ (\pi/2) = 0$.  If we did not make
this restriction, the derivative $i \hat{p} (\pi/2)$ would appear in
(\ref{eq:qh}), and when commuting with $M$ might combine with the
infinite factor $K$ to give a nonzero commutator instead of
(\ref{eq:qm-commute}).
\vspace*{0.2in}

Now let us consider conditions (a-c) above.  In each case we use a
BRST operator $Q_f$ given by (\ref{eq:ghost-BRST}) with $f_+ (\sigma)$
satisfying (\ref{eq:restriction1}, \ref{eq:restriction2}).
\vskip .1 truein

\noindent {\bf (a)} $Q_f^2 \Psi = 0$
\vspace*{0.05in}

We have
\begin{eqnarray}
\lefteqn{Q_f^2 \Psi[\phi (\sigma)] =}\label{eq:qsq} \\
 &  & 
\int_0^\pi
 d \tau\;
\int_0^\pi
 d \rho\;
f_+ (\tau) f_+ (\rho) K^2  e^{i \pi  /2}
e^{i \pi  \theta (\tau -\rho)}
e^{i \phi (\tau) + i \phi (\rho)}
\Psi\left[\phi (\sigma) +   \hskip -.05 truein \pi
\left( \theta (\tau -\sigma) + \hskip -.05\theta (\rho -\sigma) \right)
\right]\, . \nonumber
\end{eqnarray}
The factor $e^{i \pi \theta (\tau -\rho)}$ is explicitly odd under
$\rho\leftrightarrow \tau$ except at $\rho = \tau$, and the rest of
the integrand is even under this exchange, so that naively
$Q_f^2=0$. One might worry that, since the integrand contains the
divergent factor $K^2$ and does not vanish at $\rho = \tau$, a
nonvanishing result might nonetheless emerge. In Appendix A.4 we show
explicitly that this is not the case when $Q_f^2$ acts on a
well-behaved functional, such as those that lie in the ghost Fock
space.  For these states, each factor of $K$ cancels a corresponding
factor of $1/K$ when the shifted functional is expressed as a
well-behaved functional (analogous to the calculation in
(\ref{eq:inth})), and the regulated $Q_f$ is indeed nilpotent as
$x\to1$.  The question of what constitutes a well-behaved functional
is discussed at the end of this section.

We can repeat this analysis in split string language, using the
operator representation of $Q_f$ developed above.  An
identical argument shows that
\beq
\hat Q_f^2 \hat{\Psi}=0\, ,
\eeq
when $\hat{Q}_f^2$ acts on a well-behaved operator.
Consequently, using (\ref{eq:qfsss}),
\beqa
Q_f^2\Psi & \Rightarrow & \hat Q_f[\hat{Q}_f \hat{\Psi} 
-(-1)^{G_\Psi} \hat{\Psi} \hat{Q}_f ]+ (-1)^{G_\Psi} [\hat{Q}_f \hat{\Psi}
-(-1)^{G_\Psi} \hat{\Psi} \hat{Q}_f]\hat{Q}_f \nonumber \\
&=&
-(-1)^{G_\Psi} \hat Q_f  \hat{\Psi} \hat{Q}_f  +(-1)^{G_\Psi}
\hat{Q}_f \hat{\Psi}  \hat{Q}_f\\
& = & 0\,, \nonumber
\eeqa
where we have used the fact that $ \hat{Q}_f$ contains a factor of $M$
and thus has ghost number one.

While we have gone through this demonstration in some detail to be
clear about the formalism, the result that $Q_f^2 = 0$ when acting on
a well-behaved state essentially follows from the anticommutator
(\ref{eq:anticommutator1}).  In deriving this property of $Q_f$ we
have not used either of the two conditions (\ref{eq:restriction1},
\ref{eq:restriction2}) on the weight function $f_+$.
\vspace*{0.2in}

\noindent {\bf (b)} $\int Q_f \Psi = 0$
\vspace*{0.05in}

We have
\begin{eqnarray}
\int Q_f  \Psi[\phi(\s)] & = & 
\int{\cal D} \phi \; \int_0^\pi d \tau \;
e^{-\frac{3i}{2} \phi (\frac{\pi}{ 2} ) } \;
   \prod_{0\leq\s\leq{\pi\over2}}\delta\left(\phi(\s)  -  \phi(
   \pi\ -  \s)\right) \;\\
& &\hspace*{1in}\times\; \;
f_+ (\tau) K e^{i \pi/4} e^{i \phi (\tau)} \;
\Psi[\phi (\sigma) + \pi \theta (\tau -\sigma)] \nonumber
\end{eqnarray}
Performing a change of variables
$\phi (\sigma) \rightarrow \phi (\sigma) -\pi \theta (\tau -\sigma)$,
this becomes
\begin{eqnarray}
& &\int{\cal D} \phi \; \int_0^\pi d \tau \;
e^{-\frac{3i}{2} \left( \phi (\frac{\pi}{ 2} )  -\pi \theta (\tau
-\frac{\pi}{ 2} ) \right)} \;
f_+ (\tau) K e^{-i \pi/4} e^{i \phi (\tau)} \;\\
& &\hspace*{1in}
\times   \prod_{0\leq\s\leq{\pi\over2}}\delta\left(\phi(\s)  -  
\pi \theta (\tau -\sigma) -\phi(
   \pi\ -  \s)+ \pi \theta (\tau -\pi + \sigma)\right) \;
\Psi[\phi (\sigma)] \nonumber
\end{eqnarray}
%
The reason that this integral vanishes is that $f_+(\tau)$ is
symmetric under $\tau \to\pi-\tau$,
as is the overlap delta function, since
\beqa
\lefteqn{\prod_{0\leq\s\leq{\pi\over2}}\delta\left(\phi(\s) -\pi
\theta(\tau-\s) -  \phi(
   \pi\ -  \s)+\pi \theta(\tau+\s-\pi)\right)}\nonumber \\
 &=& \prod_{0\leq\s\leq{\pi\over2}} \delta\left(\phi(\s)  -
\pi \theta(\pi -\tau-\s) -  \phi( \pi\ -  \s)+\pi \theta(\s-\tau) \right)
\eeqa
while the phase factor, $\exp i[\phi(\tau)+{3\pi\over
2} \theta(\tau- {\pi\over 2})],
$ is odd. To see this note that (for $\tau< \pi/2$) the delta
function enforces:
\beq
\phi(\tau)  =\phi(\pi- \tau) -\pi \theta(\tau+\tau
-\pi)+\pi\theta(0)=\phi(\pi- \tau)+\pi/2\, ,
\eeq
since, as discussed above, $\theta(0)=1/2$.
Consequently (for $\tau< \pi/2$)
\beq
\phi(\tau)+{3\pi\over
2} \theta(\tau- {\pi\over 2}) =\phi(\tau)= \phi(\pi- \tau)+\pi/2=
\phi(\pi- \tau)+{3\pi\over
2} \theta({\pi\over 2}-\tau)-\pi \ .
\eeq
We
again need not worry about the infinite factor $K$ for well-behaved
functionals.  In this case, this is because the weight function $f_+
(\tau)$ vanishes at the point $\tau = \pi/2$, which is the one point
where the phase factor fails to be odd under $\tau \rightarrow \pi
-\tau$.  If we had chosen, for example, $f_+ (\tau) = f_-(\tau)= 1$ we
would find that the resulting operator $Q_f = c_0$ would not satisfy
$\int Q_f \Psi = 0$ even for well-behaved functionals.  For example,
this integral does not vanish when $ \Psi$ is the functional
corresponding to the Fock space state $a_1^{\dagger} | p_0 = 1/2 \rangle$.

We can repeat this analysis in the split string operator language.  We
have
\beq
\int Q_f\Psi =\Tr\left[ M^{-1}  \left(
\hat Q_f \hat\Psi-(-1)^{G_\Psi} \hat \Psi\hat Q_f\right)\right]\,  .
\label{eq:q-i-split}
\eeq
The integral $\int  Q_f\Psi$  vanishes unless the ghost number
of $\Psi$ equals two, and since
$$[\hat Q_f, M]=[\hat{\Psi}, M] =0\, ,$$
we see that (\ref{eq:q-i-split}) vanishes by cyclicity of the trace.
In general we must be careful when using the cyclicity property of the
trace, since generic operators cannot be commuted inside the trace.
For operators $\hat{\Psi}$ associated with well-behaved states this
operation is valid, however.

We have thus seen that a general ghost BRST operator of the type we
are interested in satisfies $\int Q \Psi = 0$ for well-behaved states
$\Psi$.  This result, however, depends crucially on the vanishing of
the weight function $f_+ (\pi/2)$ at the midpoint, which simplifies
the split string formalism.
\vspace*{0.2in}

\noindent {\bf (c)} $Q_f (A \star B) = (Q_f A) \star B + (-1)^{G_A}  A
\star (Q_f B)$,
\vspace*{0.05in}

We use the operator formalism  developed above,  where
\beq
Q_f(A) =\hat Q_f  \hat A
- (-1)^{G_{\Psi}}\hat A \hat Q_f\, ,
\label{eq:Qonany}
\eeq
Thus for
\begin{equation}
\quad A\star B=\hat A M\hat B\, 
\end{equation}
we have
\beq
Q_f(A\star B)= \hat Q_f \hat A M\hat B - (-1)^{G_A+G_B}\hat A
M\hat B \hat Q_f \, .
\label{eq:deriv}
\eeq
On the other hand,
\beqa
Q_f(A)\star B+(-1)^{G_A}A\star Q_f(B)&=&\left( \hat Q_f  \hat A -(-1)^{G_A} \hat A
\hat Q_f\right)
M\hat B
\nonumber
\\ && +(-1)^{G_A}\hat A
M \left( \hat Q_f  \hat B -(-1)^{G_B}
\hat B \hat Q_f\right)\, .
\label{eq:deriv2}
\eeqa
Using the fact (\ref{eq:qm-commute}) that $\hat{Q}_f$ commutes with  $M$
we see that (\ref{eq:deriv}) and (\ref{eq:deriv2}) are indeed equal.

While the derivation property of $Q_f$ follows fairly immediately in
the split string formalism, we should emphasize that this formalism,
and therefore our demonstration of this result, depended on the
vanishing midpoint condition $f_+ (\pi/2) = 0$ in the definition of
$Q$.


In this subsection we have shown that a general class of pure ghost
BRST operators satisfy the axioms of cubic string field theory when
acting on states in the bosonized ghost Fock space.  We have stated
that these results follow when the states acted on are ``sufficiently
well-behaved'', but we have not given a precise description of what
this criterion entails.  Making this statement precise amounts to
giving a precise algebraic formulation of string field theory, which
is an important outstanding problem whose resolution promises to shed
light on many questions about the theory.  In the arguments in this
section we have actually used fairly weak properties of the states
$\Psi$.  Although we have only made these results precise in the case
of states living in the bosonized ghost Fock space, for the class of
BRST operators we have considered here these statements should hold
for any functional $\Psi[\phi_n]$ which depends on $\phi_n$ for large
$n$ in the same way as the Fock space vacuum, that is as $\Psi \sim
\exp[-n\phi_n^2/2]$.  More precisely, we need to have $\Psi \sim
\exp[- \sum_{n,m}\phi_nP_{nm}\phi_m/2]$, where the eigenvalues of
$P_{nm}$ go as $n (1 +O((1/n)^\epsilon))$ for large $n$ and some
suitable value of $\epsilon$.  When working with split string
operators $\hat{\Psi}$ it seems that a slightly different criterion is
needed for a functional to be well-behaved.  Namely, the functional
$\Psi[l; r]$ should behave as $\exp[-(2k + 1)l_{2k + 1}^2/2]$ in both
the right and left degrees of freedom.  An important problem in
developing the split string field theory formalism further is the
determination of precisely how the criteria for a string state and the
corresponding string operator are related.  Some progress in
understanding the nature of well-behaved operators in the related
context of noncommutative geometry was made in \cite{Harvey-operator}.

While the subtleties of precisely which string fields are allowed in
the algebraic formulation of string field theory
will
probably become crucial in analyzing the  theory around
the D25-brane, which has a BRST operator $Q_B$ which couples the
matter and ghost sectors, it seems that the most important property we
have used here to prove (a-c) for pure ghost operators $Q$ is the
vanishing of the weight function $f_{\pm} (\pi/2) = 0$ at the
midpoint.  This allows the BRST operator to be split in a
straightforward way into right and left parts, so that the axioms of
cubic string field theory follow in a fairly straightforward fashion.
As we discuss in Section 5, this gives the vacuum string field theory
proposed by RSZ a much simpler structure than the string field theory
with BRST operator $Q_B$.

\section{Gauge Invariance and Physical Observables}

The cubic open string field theory is invariant under the gauge
transformation
\beq
\Psi \to \Psi +\delta \Psi; \quad \delta \Psi= Q(\Lambda)+\Psi \star \Lambda-
\Lambda\star \Psi  \, ,
\eeq
where $\Lambda$ is a string field of ghost number 
0 (ghost momentum $p_0 =-3/2$).  In the split string operator
formalism this variation takes the 
form
\beq
\hat \Psi \to \hat \Psi + \delta \hat \Psi; \;\;\;\;\; \quad \delta\hat  \Psi=
\hat Q_f \hat \Lambda -\hat \Lambda \hat Q_f +\hat \Psi  M \hat
\Lambda -\hat \Lambda   M
  \hat\Psi\, .
\eeq

We can write this variation in terms of a {\it covariant derivative}, 
$\hat \CD_\Psi$,
which is a functional of the string field $\hat  \Psi$.
Let us define the  covariant derivative, $\hat \CD_\Psi$, of a string 
functional $\Phi$
of ghost number zero (in the {\it fundamental} representation)  to
be
\beq
\hat \CD_\Psi \equiv\hat Q_f +  \hat \Psi M ; \quad 
\;\;\;\;\;\hat \CD_\Psi \hat\Phi
  = \hat Q_f \hat\Phi +
\hat
\Psi M
\hat\Phi \, .
\label{eq:covder}
\eeq
As discussed in Section 3, $ M = \exp[3i\phi(\pi/2)]$ goes
through  the other operators. Thus, under a gauge transformation
\beq
\hat \Psi \to \hat \Psi + \delta \hat \Psi; \;\;\;\;\;
\quad \delta\hat  \Psi=
\hat \CD_\Psi \hat \Lambda-\hat \Lambda \hat \CD_\Psi = 
\left[\hat  \CD_\Psi,\hat \Lambda\right]
\eeq
Define the field strength, $F$, to be
\beq
F=-{\p S\over \p \Psi} = Q(\Psi) +\Psi \star \Psi; \quad \;\;\;\;
\hat F =  \hat Q_f\hat\Psi +\hat\Psi \hat Q_f +\hat\Psi  M\hat\Psi\,.
\label{eq:field}
\eeq
Then it follows that under a gauge transformation,
\beq
\delta \hat F = -\hat \Lambda  M  \hat F +\hat F   M \hat 
\Lambda=-\left[\hat \Lambda  M ,
\hat F \right] , .
\label{eq:varfield}
\eeq
Clearly, if the equation of motion, $\hat F=0$, is satisfied for $\hat{\Psi}$,
it will be satisfied for $ \hat\Psi +\delta \hat \Psi$.

Note that
\beq
\hat F=M^{-1} \hat \CD_\Psi \hat \CD_\Psi = {1\over 2}M^{-1} \{\hat \CD_\Psi,
\hat \CD_\Psi\}\, ,
\eeq
The reason that the anticommutator of the covariant derivative 
arises, as opposed to the
commutator, is because $ \hat \CD_\Psi $  has ghost number one.
Also, if $\Psi$ is a solution of the equation of motion, then  $ \hat 
\CD_\Psi^2=0$,
and $ \hat \CD_\Psi $  is the BRST operator in the new vacuum defined 
by shifting the
string field  by $\Psi$.

We are interested in the largest
class of gauge invariant observables.
We believe that the most general  physical, gauge invariant observables
of open string field theory are arbitrary functions of the eigenvalues of
  $ \hat \CD_\Psi $. Under a gauge transformation  $ \hat \CD_\Psi $
changes by
\beq
   \hat \CD_\Psi  \to   \hat \CD_{\Psi+\delta \Psi} =   \hat \CD_\Psi
-\left[\hat \Lambda ,    \hat \CD_\Psi \right]
\eeq
and thus its eigenvalues are gauge invariant functionals of the 
string field $\Psi$.

What remains as a challenge is to figure out which functions of these
eigenvalues are particularly interesting observables, with interesting
physical content.  There are many such eigenvalues and one can
construct many functions of these.  In this theory, unlike ordinary
gauge theory, it is not at all obvious how to pass from the
eigenvalues of the covariant derivative to Wilson loop observables. In
ordinary gauge theory the covariant derivative carries a space-time
label and one can use it to construct local gauge invariant
observables such as $\Tr[\CD_\m,\CD_\n]^2 $. A harder case is that of
non-commutative field theory.  Consider two-dimensional noncommutative
gauge theory. In the operator formulation, as discussed in
\cite{Gross-Nekrasov}, the theory is formulated in terms of covariant
derivative operators, $\CD_\m,  (\m =1,2)$ in Hilbert space.  Although
spatial labels are now absent, one can construct the noncommutative
analogoue of Wilson loops by taking $\Tr \prod_{ i} \exp[\lambda_i^\m
\CD_{\m }]$, along a polygon generated by $\{\lambda_i^\m\}$.  In the
string field theory context, we
have neither space-time labels or space time indices for our
covariant derivative, and it is not obvious how to construct
interesting or useful observables. We would like to study something
like $\langle \Tr \;f_1( \hat \CD_\Psi ) \Tr \;f_2( \hat \CD_\Psi )
\rangle$, for appropriate functions $f_1$ and $f_2$, with the hope of
using these to probe the spectrum of the theory and to construct
scattering amplitudes. The problem is then to find the appropriate
class of $f_i$'s.

\section{Finding ghost solutions}

We now have a complete description in the split string operator
language of a class of cubic string field theories with pure ghost
BRST operators $Q$.  These theories correspond to the vacuum string
field theories proposed in \cite{rsz}.  We are interested in finding
solutions to the string field theory equations of motion
\begin{equation}
Q \Psi  + \Psi \star \Psi = 0\,.
\label{eq:SFT-equation}
\end{equation}
In \ref{sec:projection-ghost} we discuss ghost number zero projection
operators in the ghost sector.  In \ref{sec:solutions-ghost} we
describe solutions of (\ref{eq:SFT-equation}) using a pure ghost BRST
operator of the type described in \ref{sec:ghost-Q}.  We find a general
class of solutions, but find that the action vanishes generically in
the split string formalism for any solution of
(\ref{eq:SFT-equation}).  We discuss possible interpretations of this
result.  In \ref{sec:solutions-complete} we discuss related solutions
of Witten's original string field theory on the D25-brane, which has a
more complicated BRST operator $Q_B$  that couples the matter and
ghost sectors.

\subsection{Ghost projectors}
\label{sec:projection-ghost}

A first step in constructing a solution to the string field theory
equation (\ref{eq:SFT-equation}) in the ghost sector is to find a
solution of the projection equation
\begin{equation}
\chi = \chi \star \chi
\label{eq:ghost-projection}
\end{equation}
in the ghost sector.  Solutions to this equation must have ghost
number zero.  In the bosonized formulation, such solutions are of the
form
\begin{equation}
\hat{\chi} = e^{-3i \bar{\phi}/2} \hat{P}
\label{eq:ghost-projector}
\end{equation}
where $\hat{P} = | \eta \rangle \langle \eta |$ is a projection
operator on the space of midpoint-independent half-string states.  As
discussed in in \cite{rsz-3,Gross-Taylor-I}, an example of a rank one
projection operator of this form is given by the zero-momentum bosonic
sliver state constructed in
\cite{Kostelecky-Potting,Rastelli-Zwiebach}.  This state is closely
related to the D-instanton sliver discussed in Section
\ref{sec:review}.  It is useful to be slightly more explicit about
this construction, and to compare the bosonized and fermionic
representations of the ghost projector.

It was shown in \cite{Rastelli-Zwiebach} that the  sliver state can be described
in terms of Virasoro operators acting on the vacuum through
\begin{equation}
   | \Xi \rangle = \exp\left(\sum_{n = 1}^{\infty}\alpha_{2n} L_{-2n}
\right) | 0 \rangle = \exp\left( -\frac{1}{3} L_{-2} +
\frac{1}{30}L_{-4} + \cdots \right) | 0 \rangle
\label{eq:sliver-Virasoro}
\end{equation}
where $\alpha_{2n}$ are a calculable series of constants.
This formula is valid both in the ghost and matter sectors of the
theory.  The full sliver defined in conformal field theory is given by
the tensor product of the ghost and matter slivers; this state is
discussed in detail in \cite{rsz-4}.  In the matter sector, the state
(\ref{eq:sliver-Virasoro}) is defined using the zero-momentum vacuum
state $| 0 \rangle$ and the usual matter Virasoro generators
\begin{eqnarray}
L^{({\rm matt})}_n & = & \frac{1}{2} \sum_{m \neq n}
\sqrt{|m (n-m)|} \; a_m a_{n-m}+ \sqrt{n} \;a_n p_0, \;\;\;\;\; n \neq 0 \\
L^{({\rm matt})}_0 & = & \frac{1}{2} p_0^2
+ \sum_{m = 1}^{ \infty} m \;a_{-m} a_m\,. \nonumber
\end{eqnarray}
In fermionic language, the ghost sliver is defined through
(\ref{eq:sliver-Virasoro}) where the vacuum is taken to be the ghost
number zero vacuum
$| \Omega \rangle = b_{-1} | -\rangle$, and the Virasoro generators
are given by
\begin{eqnarray}
L^{({\rm g})}_n & = & (n-m) b_{n + m} c_{-m}\,.
\label{eq:fermionic-ghost-L}
\end{eqnarray}
To construct the ghost sliver in the bosonized language, we use
(\ref{eq:sliver-Virasoro}) where the ground state is given by
\begin{equation}
| \Omega \rangle  \rightarrow | p_0 = -3/2\rangle
\end{equation}
and the Virasoro generators are given by \cite{GSW,Gross-Jevicki-1}
\begin{eqnarray}
L^{(\phi)}_n & = & \frac{1}{2} \sum_{m \neq n}
\sqrt{|m (n-m)|} \; a_m a_{n-m}+ \sqrt{n} \;a_n  (p_0 -\frac{3}{2}n),
  \;\;\;\;\; n \neq 0 \\
L^{(\phi)}_0 & = & \frac{1}{2} p_0^2
+ \sum_{m = 1}^{ \infty} m \;a_{-m} a_m -\frac{1}{8} \,. \nonumber
\end{eqnarray}
These bosonized ghost Virasoro generators can be related to
(\ref{eq:fermionic-ghost-L}) using the bosonization formulae from
Section \ref{sec:bosonization}.  Thus, the fermionic and bosonized
forms of the ghost sliver defined through (\ref{eq:sliver-Virasoro})
are seen to be equivalent.  In the bosonized language, this state is
indeed a projector of the form (\ref{eq:ghost-projector}), as desired.
Once we have a single projector of this form, a variety of other
projectors can be constructed which are also of the form
(\ref{eq:ghost-projector}), with $\hat{P}$ a rank one projector on the
space of midpoint-independent half-string states, using the methods
developed in \cite{rsz-3,Gross-Taylor-I}.

\subsection{Solutions for pure ghost $Q$}
\label{sec:solutions-ghost}

Now that we have found a class of ghost number zero states satisfying
the projection equation (\ref{eq:ghost-projector}), we wish to use
this state to construct a solution to the full ghost equation
(\ref{eq:SFT-equation}).  A ghost projector (\ref{eq:ghost-projector})
can be written in operator form as
\begin{equation}
\hat{\chi} =  | \eta \rangle e^{-3i \bar{\phi}/2} \langle \eta |
  \label{eq:chi}
\end{equation}
where $| \eta \rangle$ is a half-string state such that $\hat{\chi}$
satisfies (\ref{eq:ghost-projection}).
Let us consider the state associated with the operator
\begin{equation}
\hat{\Psi} =-2 \left( \hat{Q} \hat{\chi} + \hat{\chi} \hat{Q}
\right)
\,.
\label{eq:candidate}
\end{equation}
This does not correspond to the exact state $Q \chi$, since such an
exact state corresponds to the operator
\begin{equation}
\hat{Q} \hat{\chi} -\hat{\chi} \hat{Q}\,
\end{equation}
when $\hat{\chi}$ has ghost number zero.
The equation (\ref{eq:SFT-equation})
for the state (\ref{eq:candidate})
reads (using $\hat{Q}^2 = 0$)
\begin{equation}
  \hat{Q} \hat{\chi} \hat{Q} =
\hat Q
\hat{\chi} M\hat Q \hat{\chi} +\hat{\chi} \hat Q M \hat{\chi} \hat
Q +  \hat Q \hat{\chi} M \hat{\chi} \hat Q \, ,
\end{equation}
where $M = e^{{3i\over 2}\bar{ \phi}}$ commutes with $\hat{Q}$.
This equation will be satisfied if
\beq
\hat{\chi} M \hat{\chi} = \hat{\chi}\, ;\quad \quad \hat{\chi} M\hat
Q\hat{\chi}=\hat{\chi} \hat Q M \hat{\chi} =0 \, .
\label{eq:eqofmo}
\eeq
Thus, if (\ref{eq:chi}) is a rank one projector onto a
midpoint-independent half-string state $| \eta
\rangle$ with
\begin{equation}
\langle \eta | \eta \rangle = 1
\end{equation}
and
\begin{equation}
\langle \eta | \hat{Q} | \eta \rangle = 0
\end{equation}
then we have constructed a solution to (\ref{eq:SFT-equation}).  It is
easy to see that such solutions exist for generic $Q$.  Indeed,
choosing any rank one projection, such as the state described in the
previous subsection, we can calculate
\begin{eqnarray}
\alpha_1 & = & \langle \eta | \hat{Q}_1 | \eta \rangle\,,\\
\alpha_2 & = & \langle \eta | \hat{Q}_2 | \eta \rangle\, \nonumber
\end{eqnarray}
for any pair of pure ghost BRST operators $Q_1, Q_2$.  If both
$\alpha_1 \neq 0$ and $\alpha_2\neq 0$ then we can construct a new
BRST operator
\begin{equation}
Q' = \alpha_2Q_1 -\alpha_1 Q_2
\end{equation}
which satisfies
\begin{equation}
\langle \eta | \hat{Q}' | \eta \rangle = 0\,.
\end{equation}

Thus, the expression (\ref{eq:candidate}) gives us a construction of a
wide class of solutions to (\ref{eq:SFT-equation}) for various choices
of the pure ghost BRST operator $Q$ and the ghost projector
(\ref{eq:ghost-projector}).  Combining this with a matter projector
should then give a solution to the string field theory equations of
motion in the full cubic string field theory when the BRST operator $Q$
is of the form (\ref{eq:ghost-BRST}).  According to the philosophy
expressed in \cite{rsz-2,rsz-3,Gross-Taylor-I}, this  allows us
to construct an arbitrary configuration of multiple D$p$-branes in the
vacuum string field theory postulated in \cite{rsz}.  Unfortunately,
however, there is a problem with this interpretation.  Computing the
action of a state of the form
\begin{equation}
\Psi = \Psi_m \otimes \Psi_g
\end{equation}
where $\Psi_m$ is a matter projector such as
(\ref{eq:Gaussian-projection}) and $\Psi_g$ is a ghost solution of the
form (\ref{eq:candidate}), we find that
\begin{equation}
S = -\frac{1}{6}  \int \Psi \star Q \Psi
= -\frac{4}{3}  \;{\rm Tr}\; \hat{Q} \hat{\chi} \hat{Q} \hat{\chi}
\hat{Q} = 0\,. 
\end{equation}
Thus, our solutions seem to have vanishing action\footnote{Note added:
in addition to the fact that the action vanishes for these solutions,
which as discussed in the text is generic in the split string
formalism, these particular solutions also have trivial BRST
cohomology, so they cannot be interpreted as the ghost part of a
D$p$-brane.  The vanishing of the cohomology can be easily seen by
noting that these solutions give a BRST operator $\tilde{Q} = (1-2
\hat{\chi}) \hat{Q} (1-2 \hat{\chi})$.  We would like to thank
I. Ellwood for discussions on this point.}.

This problem is actually generic to solutions in the split string
formalism.  Assume that the BRST operator $Q$ can be represented in
split string language by an operator $\hat{Q}$ so that
\begin{equation}
Q \Psi \Rightarrow \hat{Q} \hat{\Psi} -(-1)^{G_\Psi} \hat{\Psi}
\hat{Q}
\end{equation}
where $\hat{Q}$ has the properties we derived for operators of the
form (\ref{eq:ghost-BRST}) in Section \ref{sec:ghosts}, namely
$\hat{Q}^2 = 0$ and $ {\rm Tr}\; \hat{Q} \hat{\Psi} = {\rm Tr}\;
\hat{\Psi}\hat{Q}$.  Then for any solution of the equation
\begin{equation}
Q \Psi = \Psi \star \Psi
\end{equation}
we have
\begin{eqnarray}
\int  \Psi \star Q \Psi & \Rightarrow &
{\rm Tr}\; \hat{\Psi} (\hat{Q} \hat{\Psi} + \hat{\Psi} \hat{Q})
\nonumber\\
  & = &2  \;{\rm Tr}\; \hat{Q} \hat{\Psi} \hat{\Psi} \nonumber\\
  & = &-2  \;{\rm Tr}\; M^{-1} \hat{Q}(\hat{Q} \hat{\Psi} + \hat{\Psi}
\hat{Q}) \nonumber\\
  & = &-2 \; {\rm Tr}\; M^{-1} \hat{Q} \hat{\Psi} \hat{Q} \nonumber\\
& = & 0\,. \label{eq:vanishing}
\end{eqnarray}
In this calculation
we have only used the above-mentioned properties of the BRST operator
$Q$ and the cyclicity of the trace.  All these properties were
explicitly verified in Section \ref{sec:ghosts} for the simple class
of pure ghost BRST operators which are linear in $c$ with a weight
which vanishes at the midpoint $\sigma = \pi/2$, when dealing with
string fields $\Psi$ which are in the string Hilbert space, or which
are similarly well-behaved.

There are a number of possible explanations for the vanishing of the
action (\ref{eq:vanishing}).  We list here some of the most
likely possibilities:
\vspace*{0.2in}

\noindent
{\bf (A)} One possibility is that the solution of the vacuum string
field theory corresponding to a D$p$-brane lies outside the Hilbert
space.  A crucial question, to which we do not yet know the answer, is
precisely what the space of allowable string field functionals should
be.  It seems that this space is larger than the string Hilbert space.
Examples of states which probably lie outside the Hilbert space are
given by the matter sliver state discussed in Appendix
\ref{sec:a-projector} and the matter state $| 0 \rangle \star | 0
\rangle$.  The sliver state lies outside the Hilbert space of
normalizable Fock space states when the matrix $S$ given in
(\ref{eq:s}) has an eigenvalue of $\pm 1$.  There is some evidence
\cite{rsz-4,Sen-personal} that indeed $S$ has an eigenvalue of $-1$, taking
the sliver state just outside the Hilbert space.  This in turn
suggests that the matrix $V$ from (\ref{eq:v}) has an eigenvalue of
$-1$ in the odd-odd sector, so that the state $|0 \rangle \star |0
\rangle$ also lies just outside the normalizable Fock space.  In order
to have a solution of (\ref{eq:SFT-equation}) which has nonvanishing
action in the cubic string field theory with a pure ghost BRST
operator of the form (\ref{eq:ghost-BRST}), it may be that we need to
understand how the properties of $Q$ derived in \ref{sec:ghost-Q} need
to be modified for states in a larger class of functionals.  We must
be careful, however, not to expand the space of allowed functionals
too far, or the structure of the theory, which depends critically on
the properties of $Q$, will break down.  One piece of evidence which
may support this possible explanation of (\ref{eq:vanishing}) is the
recent numerical calculation of a ghost solution with nonvanishing
action in a level truncation approximation around the ghost sliver in
the vacuum string field theory \cite{David-sliver}.  At least to level
2 truncation, this solution appears to be numerically stable, and has
a structure roughly compatible with the ansatz (\ref{eq:candidate}).
This may suggest that the action does not actually vanish for all
solutions, and that the solutions of interest indeed lie just far
enough outside the Fock space that the steps used in
(\ref{eq:vanishing}) are not exactly valid, but that the solution is still a
well-enough behaved functional that the structure of the string field
theory still holds.
\vspace*{0.15in}

\noindent
{\bf (B)} Another possible explanation for the vanishing of the action
(\ref{eq:vanishing}) for a generic solution of the theory with a pure
ghost BRST operator $Q$ is that the vacuum string field theory
postulated in \cite{rsz} may have an action which is related by an
infinite overall normalization factor to the action of Witten's
original theory in the condensed tachyon vacuum.  Since there are no
perturbative physical states in the stable vacuum
\cite{Sen-universality,Ellwood-Taylor}, there is no simple way of
calculating the overall normalization of the action in the vacuum
string field theory other than by comparing the energy of soliton
configurations such as those considered here.  We describe in Appendix
\ref{sec:a-projector} a set of normalization factors which are needed
to convert from the split string formalism to the usual Fock space
conventions, in the matter sector.  If as mentioned above, the matrix
$V$ has an eigenvalue $-1$ in the odd-odd block, the product of
conversion factors $\gamma_1\gamma_3$ vanishes.  It may be that
similar formally vanishing or infinite normalization factors are
needed to establish the finite energy of a D$p$-brane solution in the
vacuum string field theory.
\vspace*{0.15in}

\noindent
{\bf (C)} A final possible explanation for the vanishing of
(\ref{eq:vanishing}) may be that the vacuum string field theory
hypothesis, in which the BRST operator $Q$ is taken to be linear in
the ghost field and vanishing at the string midpoint, may simplify the
theory too much.  Indeed, the derivation of the properties of the BRST
operator $Q$, such as $\int Q \Psi = 0$, in Section \ref{sec:ghosts}
depended crucially on the vanishing of the weight function $f
(\sigma)$ at the string midpoint $\sigma = \pi/2$.  In the original
string field theory of Witten with the BRST operator $Q_B$ coupling
the matter and ghost sectors, the condition that $\int Q_B \Psi = 0$
depends on anomaly cancellation between the matter and ghost sectors,
which occurs only in dimension 26 \cite{Gross-Jevicki-2}.  It may be
that by choosing a particularly simple class of BRST operators for
which the matter and ghost sectors decouple and for which the anomaly
plays less of a role, we have lost some important structure in the
theory, so that the action will indeed generically vanish for
solutions of the string field theory equations.  It seems likely that
the theory in this case retains something like the topological
structure of the original theory (for example the existence of D-brane
solutions), but perhaps not all the dynamical details.  One issue
which may be related to this possibility is the question of how to
find poles at nonzero $p^2$ in the theory around the D$p$-brane
solitons in the vacuum string field theory.  Some discussion of
questions related to this issue was given in \cite{rsz-4}, but it is
still unclear how the physical spectrum of the usual open string can
be constructed using a $Q$ with no dependence on the matter sector.
\vspace*{0.15in}

We conclude this discussion of solutions of the equations of motion in
the case of pure ghost BRST operators with a few remarks on how these
solutions might be checked using level truncation.  The split string
operator $\hat{Q}$ given in (\ref{eq:qh}) can be described in terms
of operators $Q_l, Q_r$ acting on the space of full string functionals
through
\begin{eqnarray}
Q_l \Psi & \Rightarrow &  \hat{Q} \hat{\Psi}\\
Q_r \Psi & \Rightarrow &   \hat{\Psi}\hat{Q}\nonumber\,.
\end{eqnarray}
{}From the definition (\ref{eq:ghost-BRST}), we have
\begin{eqnarray}
Q_l &= & \frac{1}{2 \pi} \int_0^{\pi/2} d \sigma \; \left(f_+ (\sigma)
c^+ (\sigma)
+ f_-(\sigma) c^-(\sigma) \right) \label{eq:qlr}\\
Q_r &= & \frac{1}{2 \pi}
\int_{\pi/2}^\pi d \sigma \; \left(f_+ (\sigma) c^+ (\sigma)
+ f_-(\sigma) c^-(\sigma) \right) \nonumber\,.
\end{eqnarray}
This gives us a way of explicitly defining $Q_l, Q_r$ in terms of
fermionic ghost raising and lowering operators.
For example, when $f_+ = f_-= 1$, we have
\begin{equation}
Q = c_0,
\end{equation}
and
\begin{eqnarray}
Q_l & = & \frac{1}{2} c_0 +
\sum_{k} \frac{(-1)^k}{ \pi (2k + 1)} c_{2k + 1}  \\
Q_r & = & \frac{1}{2} c_0   -
\sum_{k} \frac{(-1)^k}{ \pi (2k + 1)} c_{2k + 1}  \,. \nonumber
\end{eqnarray}
In this language the solution (\ref{eq:candidate}) is given by
\begin{equation}
| \Phi \rangle = 2 (Q_r-Q_l) |\chi \rangle
\label{eq:Fock-candidate}
\end{equation}
where $| \chi \rangle$ is the ghost Fock space state associated with a
projection operator satisfying $| \chi \rangle \star | \chi \rangle =
| \chi \rangle$.  This choice of $Q$ does not satisfy all the
desired axioms, since $f (\pi/2) \neq 0$, but the same approach gives
the analogous operators $Q_l, Q_r$ for any other choice of $f$.  It
would be interesting to check using level-truncation whether states of
the form (\ref{eq:Fock-candidate}) indeed correspond with solutions of
the vacuum string field theory equations of motion having nonzero
action.  Some preliminary numerical work towards finding ghost
solutions appeared in
\cite{David-sliver}; the results of this work seem very roughly compatible
with this solution ansatz (for example, the two states considered
there with a single $c$ acting on the sliver are weighted much more
heavily than the single state with a $bcc$ product acting on the
sliver in a 3-state truncation), although more data is needed to make
any definitive statement about this question.

\subsection{Solutions of the coupled matter-ghost theory}
\label{sec:solutions-complete}

It is clearly important to understand how the approach we have taken
in this work can be generalized to the perturbative string field
theory around a D25-brane given by Witten's original formulation of
the theory with BRST operator $Q_B$.  All of the formal structure we
have developed here for describing the string star product $\star$ and
integral $\int$ in the matter and ghost sectors applies equally well
to the D25-brane theory.  The difficult remaining technical problem is
to formulate the split string description of $Q_B$ and to check that
it satisfies the desired axioms.  We leave the details of this
formulation to future work, but make some general comments here
outlining some possible issues which may arise in solving this
problem.

The BRST operator $Q_B$ is given by
\begin{equation}
Q_B = \int_0^\pi d \sigma \; \left[ c_+ (\sigma) \left(T_{--}^{(m)} (\sigma)
+\frac{1}{2}T_{--}^{(g)} (\sigma)
\right)+ \left(+ \leftrightarrow -\right)\right]\,.
\end{equation}
This operator can be rewritten in a fairly straightforward fashion in
terms of the bosonized ghosts in a form similar to
(\ref{eq:ghost-BRST}), where $f$ is replaced by expressions of the
form $\partial x \partial x$ in the matter sector and $\partial \phi
\partial \phi + \partial^2 \phi$ in the ghost sector.  This operator
can in principle be split into parts acting on the left and right of
the string along the lines of (\ref{eq:qfsss}, \ref{eq:qlr}).  This
splitting is more subtle, however, than the splitting in the case of a
pure ghost operator with weight vanishing at the midpoint.  The
verification of the axioms of string field theory using the split form
of $Q_B$ is also much more subtle.  For example, as demonstrated in
\cite{Gross-Jevicki-1}, the proof that $\int Q_B \Psi= 0$ depends
crucially on anomaly cancellation between the matter and ghost
sectors.

Despite these complications, we believe that a split string
formulation of the BRST operator $Q_B$ is possible.  It may be,
however, that the action of $Q_B$ on even a well-behaved state is more
complicated than (\ref{eq:qfsss}), due to the midpoint anomaly.  If,
indeed, the structure of the split BRST operator is more complicated
in the case of $Q_B$ than in the pure ghost case, it may be possible
to avoid the vanishing action problem of (\ref{eq:vanishing}), even
for well-behaved string states.  This would fit well with Sen's
conjectures \cite{Sen-universality} and results from
level truncation indicating the existence of nontrivial solutions with
nonzero action \cite{Kostelecky-Samuel,Sen-Zwiebach,Moeller-Taylor}.

It may even be that the solution corresponding to the locally stable
vacuum in the original D25-brane string field theory may have
something like the form of (\ref{eq:candidate}), and that subtleties related
to the midpoint anomaly may be responsible for giving this solution a
nonvanishing action.  It is worth pointing out that the solution
(\ref{eq:candidate}) is closely related to a solution $Q_l  | I
\rangle$ which was considered some time ago in \cite{cubic}, and
which leads to a purely cubic string field theory action.
Indeed, (\ref{eq:candidate}) is equivalent to this solution when the
rank one projection $\hat{\chi}$ is replaced by the infinite rank
projection $\hat{I}$.  It may be that the solutions
(\ref{eq:candidate}) with $\hat{\chi}$ a rank one projector correspond
to single D-branes, while the solution with $\hat{\chi} = I$ may
represent a $U (\infty)$ theory of an infinite number of space-filling
D25-branes, such as is useful in K-theory constructions
\cite{Witten-K}.  In any case, we believe that the technology we have
developed here provides a good basis for further development towards
finding an analytic solution of the full string field theory with BRST
operator $Q_B$, as well as a simple formulation of this theory using
the split string approach.

\section{Conclusions}

In this paper we have continued the work begun in
\cite{Gross-Taylor-I} of developing a complete description of
string field theory in terms of an algebra of operators on half-string
states.  In \cite{Gross-Taylor-I} we developed this formalism in the
matter sector of the theory and identified certain states of interest
as projection operators in the matter sector.  In this paper we have
developed the split string formalism for the string field star product
and integral in the ghost sector, using the bosonized description of
the ghosts as a single bosonic scalar field.  We also described a
class of pure ghost BRST operators $Q$ in the split string language.
These operators were conjectured in \cite{rsz} to define a cubic
string field theory in the closed string vacuum which is formally a
field theory of open strings but which has no physical open string
states.  This vacuum string field theory shares with the earlier
proposal of a purely cubic string field theory \cite{cubic} the virtue
of being independent of a choice of conformal field theory background
\cite{rsz-4}.

We used the split string formalism for the ghost sector to find a
class of solutions to the string field theory equations of motion.  We
showed that these solutions all have vanishing action, and that in
fact the split string formalism for the pure ghost $Q$'s considered
here gives vanishing action for any solution which is described by a
well-behaved functional, such as a state in the Hilbert space.  We
speculated that this generically vanishing action might be reconciled
with the expectation that D-brane states have finite nonzero action in
several ways.  It may be that the solutions of interest lie far enough
outside the usual string Hilbert space that the arguments used here do
not apply.  It may be that the action of the vacuum string
field theory is related to that of the usual cubic string field theory
with BRST operator $Q_B$ by a formally infinite multiplicative factor,
leading to a vanishing action for D-branes in the vacuum string field
theory.  It may also be that the simplifications arising from the
assumption that the BRST operator have a pure ghost form and have
vanishing action at the string midpoint leads to a theory with less
structure than the original theory with BRST operator $Q_B$, so that
in the theory with pure ghost $Q$ D-brane solutions still exist but
have vanishing action.  One example of such a simplification is the
absence of anomalies associated with the action of pure ghost $Q$'s at
the midpoint, suggesting that the classical vacuum string field theory
of RSZ can be consistently defined in any dimension, while the theory
with BRST operator $Q_B$ only works in dimension 26. 

In this paper we also discussed how the proposed solutions to the
ghost equations described here could be studied in level truncation,
and we used the solutions in the case of pure ghost $Q$ to
suggest a form for solutions to the original D25-brane theory with
BRST operator $Q_B$.  This class of solutions is related to the
solution suggested in \cite{cubic} which takes the original theory to
a purely cubic action, but in our solutions the string identity
operator is replaced by a finite rank projector in the matter and
ghost sectors.

The work we have presented here represents only one  step towards
the complete realization of open string field theory as a
background-independent string field theory of operator algebras which
will hopefully eventually describe closed strings as well as open
strings.  There are many major questions which must be answered before
this program can be completed.  First, we need to find a
representation of the usual perturbative string BRST operator $Q_B$
which couples the matter and ghost sectors in terms of operators on
the space of half-string states.  Next, we need to find explicit
analytic solutions of the resulting string field equations of motion,
possibly using the form of solution suggested here.  After such
solutions are found, we need to show that these solutions can be
identified with multiple D$p$-brane configurations, including the
identification of the classical vacuum state which has so far only
been understood using level-truncation methods.  Once the solution
space of the theory is understood, we need to show analytically that
in the closed string vacuum there are no perturbative open string
excitations and that around the various multiple D$p$-brane solutions
we have the expected spectrum of open string states.

We are optimistic that the problems just listed represent a series of
incremental improvements which can be realized by further development
of the formalism described here or by related methods.  There are also
some larger questions whose solution will probably involve substantial
new conceptual developments in the theory.  These questions involve
the identification of asymptotic closed string states in the open
string field theory, the determination of the relevant class of
physical observables for the theory, the abstract formulation of the
theory as a theory of operator algebras with no reference to any
background geometry, and the generalization of all this structure to
the superstring.  If substantial progress can be made on these
problems, open string field theory has the potential to become the
first truly background-independent nonperturbative description of
string theory.

\appendix

\section{Appendices}

\subsection{Relative normalizations using Fock space states
and functionals}
\label{sec:a-normalization}

In this appendix we describe some details of the connection between
the split string description of string field theory we are developing
here and the more traditional Fock space approach of
\cite{Gross-Jevicki-1,Gross-Jevicki-2}.  In particular
there are a number of normalization factors which are different in the
two approaches, and we want to be able to translate back and forth
between the Fock space and split string languages at will.  In this
appendix we restrict attention to the matter sector and compare
normalization factors between the Fock space and split string
pictures.  In Appendix \ref{sec:a-projector} we show explicitly
how these normalization factors are related in the split string and
Fock space descriptions of a particular rank one projection operator
considered in \cite{rsz-3,Gross-Taylor-I}.

In the Fock space language, the basic elements of cubic string field
theory can be defined in terms of three states $\langle I |, \langle
V_2 |,$ and $\langle V_3 |$ which lie respectively in the 1-fold,
2-fold, and 3-fold tensor products of the dual string Hilbert space.
These three states are calculated in \cite{Gross-Jevicki-1}, using a
particular choice of normalization of the states.  We compare those
normalizations here to those which arise in the split string approach.
To make these comparisons, we define the normalized functional
associated with the Fock
space vacuum $| 0 \rangle$ to be
\begin{equation}
  \Psi_0 = k\exp \left(-x_0^2 -\frac{n}{2}x_n^2 \right)
\end{equation}
where
\begin{equation}
k = \left(\frac{2}{\pi} \right)^{26/4}
\prod_{n = 1}^{\infty}\left( \frac{n}{\pi}  \right)^{26/4}
 \label{eq:k}
\end{equation}

We now compare the normalization factors between the basic string
field theory elements $\int$ and $\star$ as defined by
(\ref{eq:integral-full}, \ref{eq:integral-star}) and as defined by
$\langle I |, \langle V_2 |$ and $\langle V_3 |$.
The identity state $\langle I |$ is defined so that
\begin{equation}
\int \Psi = \gamma_1
\langle I | \Psi \rangle
\end{equation}
where $\gamma_1$ is a numerical constant.  In \cite{Gross-Jevicki-1},
$\langle I |$ is defined through
\begin{equation}
\langle I | = \langle 0 | \exp \left[ -\frac{1}{2}
  \left( a | C | a \right) \right]
\end{equation}
where
\begin{equation}
C_{nm} = (-1)^n \delta_{nm}\,.
\end{equation}
to calculate the numerical constant $\gamma_1$, we compute the integral
of the vacuum
\begin{eqnarray}
\int \Psi_0 & = &
  \int \; \prod_{n = 0}^{\infty}
dx_n \;\prod_{k = 0}^{ \infty}  \delta (x_{2k + 1}) \;
\Psi_0[\{x_n\}]
  \\
  & = &  \left(2\pi\right)^{26/4} \left(
 \frac{1}{\pi}  \cdot \frac{4\pi}{2}
\cdot \frac{3}{ \pi} \cdot
\frac{4 \pi}{ 4}   \cdots \right)^{26/4}\,.
\end{eqnarray}
Since
\begin{equation}
\langle I | 0 \rangle = 1\,,
\end{equation}
we have
\begin{equation}
\gamma_1 =  \left(2\pi\right)^{26/4} \left(
 \frac{1}{\pi}  \cdot \frac{4\pi}{2}
\cdot \frac{3}{ \pi} \cdot
\frac{4 \pi}{ 4}   \cdots \right)^{26/4}\,.
\end{equation}
We will find it useful to leave this expression in the form of an
infinite product, although it could be regulated in various ways.

Now let us consider the two-string vertex $\langle V_2 |$.  This
vertex satisfies
\begin{equation}
\int \Psi_1 \star \Psi_2 = \gamma_2 \langle V_2 |
\left( | \Psi_1 \rangle \otimes | \Psi_2 \rangle \right)
\end{equation}
for another constant $\gamma_2$.  $\langle V_2 |$ is determined in
\cite{Gross-Jevicki-1} to be given by
\begin{equation}
\langle V_2 | = (\langle 0| \otimes \langle 0 |)
\exp \left[ -\frac{1}{2}\left( a_{(1)} | C | a_{(2)}\right) \right]
\end{equation}
where $a_{(1, 2)}$ are the lowering operators in the first and second
Hilbert spaces respectively.  It is straightforward to compute
\begin{equation}
\int \Psi_0 \star \Psi_0 =
\langle V_2 | (| 0 \rangle \otimes | 0 \rangle) = 1\,
\end{equation}
so that we have
\begin{equation}
\gamma_2 = 1\,.
\end{equation}

We now move on to the three-string vertex  $\langle V_3 |$.  This
vertex satisfies
\begin{equation}
\int \Psi_1 \star \Psi_2 \star \Psi_3
= \gamma_3 \langle V_3 |
\left( | \Psi_1 \rangle \otimes | \Psi_2 \rangle
\otimes | \Psi_3 \rangle \right)
\end{equation}
for another constant $\gamma_3$.  $\langle V_3 |$ is determined in
\cite{Gross-Jevicki-1} to be given by
\begin{equation}
\langle V_3 | = (\langle 0| \otimes \langle 0 |\otimes \langle 0 |)
\exp \left[-\frac{1}{2}
 \left( \sum_{i, j = 1}^3  a_{(i)} | V^{ij} | a_{(j)}\right) \right]
 \label{eq:v}
\end{equation}
for constants $V^{ij}_{nm}$ which are explicitly given in
\cite{Gross-Jevicki-1,cst,Samuel,Ohta}.  The three-string vertex can be
used to calculate the star product of a pair of states through
\begin{equation}
\langle \Psi_1 \star_F \Psi_2 | =
\langle V_3 |
\left( | \Psi_1 \rangle \otimes | \Psi_2 \rangle
\otimes  \cdot \right)\,,
  \label{eq:Fock-star}
\end{equation}
where by $\star_F$ we distinguish the Fock space star product from
that defined through (\ref{eq:matter-star}, \ref{eq:half-star}) which
has a normalization differing by $\gamma_3$,
\begin{equation}
\Psi_1 \star \Psi_2 = \gamma_3 \left(  \Psi_1 \star_F \Psi_2 \right)\,.
\end{equation}
Using (\ref{eq:Fock-star}), we can
compare
\begin{equation}
\int \Psi_0 \star \Psi_0 = 1
\end{equation}
with
\begin{eqnarray}
\langle I | \left( | 0 \rangle \star_F | 0 \rangle \right) & = &
\langle 0 |\exp  \left[-\frac{1}{2}\left( a | C | a\right) \right]
\exp \left[ -\frac{1}{2} \left(a^{\dagger} | V | a^{\dagger}\right)
\right] | 0 \rangle\\
  & = & \frac{1}{ [\sqrt{\det (1-CV)}]^{26}}
\end{eqnarray}
where $V =V^{11}$, to find that
\begin{equation}
\gamma_3 = \frac{ 1}{ \gamma_1} [\sqrt{\det (1-CV)}]^{26}\,.
\label{eq:gamma-3}
\end{equation}
We note in passing that $V_{nm} = 0$ unless $n \equiv m \; ({\rm mod}
\; 2)$, so that $V$ can be decomposed into even and odd blocks $V_e,
V_o$.  We use this fact in the following appendix.

In this appendix we have described the relative normalization
factors used in the Fock space and split string approaches to string
field theory, in the matter sector.  One particular advantage of the
normalization conventions natural to the split string framework is
that rank one projection operators satisfying
(\ref{eq:matter-projection}) naturally have a unit trace
\begin{equation}
\int \Psi = {\rm Tr}\; \hat{\Psi} = 1\,.
\end{equation}
In the following  appendix we discuss the connection with the
analogous statement in the Fock space description.

\subsection{Normalization of rank one projectors}
\label{sec:a-projector}

In \cite{Kostelecky-Potting,rsz-2}, a string field $| \Phi_F \rangle$ was
constructed using Fock space methods which satisfies the projection equation
\begin{equation}
| \Phi_F \rangle =| \Phi_F \rangle\star_F | \Phi_F \rangle\,
\label{eq:Fock-projector}
\end{equation}
defined using the Fock space normalization for the star product $\star_F$
in (\ref{eq:Fock-star}).
This state is given by
\begin{equation}
| \Phi_F \rangle
=[\sqrt{\det (1-Z)\det (1 + T)}]^{26}
\exp \left[ -\frac{1}{2}
\left( a^{\dagger} | S | a^{\dagger} \right) \right]
| 0 \rangle
\label{eq:example-projector}
\end{equation}
where
\begin{equation}
Z = CV,
\end{equation}
\begin{equation}
S = CT,
 \label{eq:s}
\end{equation}
and
\begin{equation}
T = \frac{1}{2Z}
\left( 1 + Z-\sqrt{(1 + 3Z) (1-Z)} \right)\,.
\end{equation}
Note that like $V$,  $S$ is only nonzero in the even-even and odd-odd
blocks $S_e, S_o$.
The sliver state (\ref{eq:example-projector}) is discussed further in
\cite{rsz-4,Matsuo}.

In \cite{Gross-Taylor-I}, it was shown that the state
(\ref{eq:example-projector}) is a rank one projector on the space of
half string functionals by demonstrating that the dependence of this
state on the left and right modes is described by a functional of the
form
\begin{equation}
\exp \left( -\frac{1}{2}\, l \cdot M \cdot l-\frac{1}{2}\,r \cdot M \cdot
r \right)
\label{eq:m-decomposition}
\end{equation}
where $M$ is given by
\begin{equation}
M =\frac{1}{2} E_o^{-1} \frac{1-S_o}{1+S_o}  E_o^{-1}\,.
\label{eq:m}
\end{equation}
In this expression, $M_{2j + 1, 2k + 1}$ and all other matrices are
restricted to the odd-odd block.

We now use the relationship between
the Fock space and split string normalization conventions derived
above to demonstrate that the state proportional to
(\ref{eq:example-projector}) which satisfies the split string
projection equation indeed has the normalization
(\ref{eq:Gaussian-projection}).  This serves as a check on our
formalism and completes the analytic  construction of a rank one
projector satisfying $\hat{\Phi} = \hat{\Phi} \hat{\Phi}$ and ${\rm
Tr}\; \hat{\Phi} = 1$ in the split string formalism.

 From the discussion above of the difference between the Fock space and
split string normalization of the star product, we see that associated
to a state $| \Phi_F \rangle$ satisfying (\ref{eq:Fock-projector}) there
should be a functional
\begin{equation}
\Phi = \frac{1}{ \gamma_3}  \Phi_F
\label{eq:correct-functional}
\end{equation}
which satisfies
\begin{equation}
\Phi = \Phi \star \Phi
\end{equation}
(using the functional integral normalization of the star product
(\ref{eq:integral-star})).  The Fock space state
(\ref{eq:example-projector}) has the functional description
\begin{equation}
\Phi_F[x] =
[\sqrt{\det (1-Z)\det (1 + T)}]^{26}
\; \frac{k}{[\sqrt{\det 1 + S}]^{26}}  \;
\exp \left[
-\frac{1}{2}\left( x | L | x \right) \right]
\end{equation}
where $k$ is given in (\ref{eq:k}) and
\begin{equation}
L = E^{-1} \left( \frac{1-S}{ 1 + S}  \right) E^{-1}\, .
\end{equation}
In \cite{Gross-Taylor-I}, we showed that the exponential factor
decomposes into a product of Gaussians of the left and right modes of
the form (\ref{eq:m-decomposition}).  To show that $\Phi$ given by
(\ref{eq:correct-functional}) is indeed of the form
(\ref{eq:Gaussian-projection}), it remains to show that the
normalization factors match, that is that
\begin{equation}
\left(\det  \frac{M}{\pi}  \right)^{26/2}
= \frac{k}{\gamma_3}
[\sqrt{\det (1-Z)\det (1 + T)}]^{26}
\; \frac{1}{[\sqrt{\det 1 + S}]^{26}} \,.
\label{eq:checking}
\end{equation}
 From (\ref{eq:gamma-3}) and the definition of $Z = CV$, this reduces to
\begin{equation}
\left(\det  \frac{M}{\pi}  \right)^{26/2}
= k \gamma_1
\; \frac{[\sqrt{\det (1 + T)}]^{26}}{[\sqrt{\det 1 + S}]^{26}} \,.
\end{equation}
Breaking $S, T$ into odd and even blocks we have
\begin{eqnarray}
\det (1 + S) & = &  \det (1 + S_o) \det (1 + S_e)  \nonumber\\
\det (1 + T) & = &  \det (1 - S_o) \det (1 + S_e)\,.
\end{eqnarray}
 From (\ref{eq:m}) we have
\begin{equation}
\det M = \det\left[ \frac{E_o^{-2}}{2}  \right]
\frac{\det (1 -S_o)}{ \det (1 + S_o)} \,.
\end{equation}
It follows that all the determinants of $1 \pm S_{o, e}$ cancel in
(\ref{eq:checking}), so it remains to verify that the constant factors
cancel, which occurs when
\begin{equation}
\det \left( \frac{E_o^{-2}}{2 \pi} \right)  = k \gamma_1\,
\end{equation}
which is equivalent to
\begin{equation}
\prod_{k = 0}^{\infty}
\left( \frac{2k + 1}{2 \pi}  \right)^{26/2}
= \prod_{k = 0}^{\infty}
\left[\left( \frac{2k}{\pi}   \cdot\frac{2k + 1}{\pi}  \right)
\left( \frac{4 \pi}{2k} \cdot  \frac{2k + 1}{ \pi}  \right) \right]^{26/4}.
\end{equation}
which is manifestly true for each $k$.  
This completes our check that the state (\ref{eq:correct-functional})
is a rank one half-string projection satisfying
\begin{equation}
\Phi = \Phi \star \Phi
\end{equation}
  and
\begin{equation}
\int \Phi = 1.
\end{equation}

\subsection{Integrals}

Here we evaluate an integral used in Section 3.1.2.

We wish to evaluate
\begin{equation}
\int_{-\pi}^\pi {d \sigma \over 2 \pi}
{e^{-i{\s\over 2}+i \e(\s)\pi/4 +i n\s}\over   (2-2 \cos 2
\sigma)^{1/4}}\,, \;\;\;\;\; n \geq 0\,.
\end{equation}
For $n\geq 0$, we have
\beq
 \int_0^\pi {d\s\over 2 \pi} {e^{\pm
i({\s\over 2}+ i{n\s
})}\over
\sqrt{2\sin(\s)}} ={\sqrt{\pm i}\over 2 \pi}\int_{-1}^1 {dx}{x^n\over
\sqrt{1-x^2}}={\sqrt{\pm i}\over 2^{n+1 }  }  {n \choose
n/2}{1+(-1)^n\over 2}\, ,
\eeq
where we changed variables according to $\s\to x=\exp(\pm i\s)$.
Similarly, for $n > 0$ we have
\beq
   \int_0^\pi {d\s\over 2 \pi}
{e^{\pm i({\s\over 2}-i{n\s
})}\over
\sqrt{2\sin(\s)}} = {\sqrt{\mp i}\over
2^{n }  }  {n-1 \choose (n-1)/2}{1+(-1)^{n+1}\over 2 }\,  .
\eeq
Using these we can easily establish that (for $n\geq 0$)

\beq
\int_{-\pi}^\pi {d \sigma \over 2 \pi}
{e^{-i{\s\over 2}+i \e(\s)\pi/4 +i n\s}\over   (2-2 \cos 2
\sigma)^{1/4}}= \delta_{n,0}\, .
\label{eq:integral}
\eeq
(For $n<0$ this integral does not vanish).
Consequently the integral within the square bracket in
(\ref{eq:inth}) is identically one
and independent of the $\phi_n$'s.
Another way to see this result is to note that
\begin{equation}
{e^{-i{\s\over 2}+i \e(\s)\pi/4}\over   (2-2 \cos 2
\sigma)^{1/4}}=\exp \left(
-i \sigma/2 + i\e (\sigma) \pi/4 + \sum_{m= 1}^{\infty} \frac{\cos 2m
\sigma}{
2m}\right)  =\exp \left(
  \sum_{m = 1}^{\infty}\frac{e^{2im \sigma}}{ 2m}  \right)
\end{equation}
so that (\ref{eq:integral}) follows immediately.

\subsection {$Q^2=0$}

In the proof that  $Q^2=0$, we ignored the fact that the formally vanishing
double integral in (\ref{eq:qsq}) had a nonzero integrand at $\rho =
\tau$, which is multiplied by the divergent factor $K$, so that a finite result
might in principle be obtained.
If  we use the regulated form of $Q_f$ the factor
$K^2e^{i\pi \t(\tau-\rho)}$, which is odd under $\rho \leftrightarrow
\tau$ is replaced by
$$
K_x^2\exp i\left[{\pi \over 2} + {1\over 2i} \ln{\e +i(\tau -\rho)
\over \e -i(\tau -\rho)}\right] \, .
$$
In the integral over $\rho$ and $\tau $ in (\ref{eq:qsq}), we can
symmetrize in $\tau$ and $\rho$, so that
this factor becomes (recall that $K_x \sim  1/\sqrt{1-x}$ and $x=1-\e$),
\beqa
{i\over \e} \left[\sqrt{\e +i(\tau-\rho) \over \e-i(\tau-\rho)} +
\sqrt{\e -i(\tau-\rho) \over \e+i(\tau-\rho)} \right]
& = &{2i \over \sqrt{(\tau-\rho)^2+\e^2}}
\, .
\eeqa
Therefore, due to the divergent factor of $K$, the formally vanishing
$K^2\left(e^{i\pi \t(\tau-\rho)}+e^{i\pi \t(\rho-\tau)}\right)$
can yield a finite result. However, this does not invalidate the fact
that $Q^2 \Psi=0$ for
well-behaved functionals $\Psi$.
The essential point is that in (\ref{eq:qsq})
we encounter
$$\exp(i\phi(\tau)+i\phi(\rho))) \;\Psi\left[\phi (\sigma) +    \pi
\left( \theta (\tau -\sigma) + \theta (\rho -\sigma) \right)
\right],$$
a string field to which two kinks have been added.
If $\Psi$ is a nice functional, such as a Fock space state of the form
$P(\phi_n)\Psi_0[\phi (\sigma)]$,
where $P(\phi_n)$  is a polynomial
in the $ {\phi_n}$'s and  $\Psi_0$ is the vacuum functional
\begin{equation}
\Psi_0[\phi (\sigma)] =
C e^{-i \phi_0/ 2} \exp \left( -  \sum_{n = 1}^{ \infty}
   \frac{n}{2} \phi_n^2 \right)\, .
\end{equation}
then the shifted vacuum functional is of the form
\beqa
\lefteqn{
\exp(i\phi(\tau)+i\phi(\rho))\Psi_0\left[\phi (\sigma) +   \pi
\left( \theta (\tau -\sigma) + \theta (\rho -\sigma) \right)
\right] }   \nonumber \\ &=& 
C
   e^{i( {3\phi_0-\tau-\rho)/ 2}} \exp \left(   i\sqrt{2}\sum_{n = 1}^{ \infty}
   \phi_n(\cos(n\tau)+\cos(n\rho))  \right.\\
& &\hspace*{2in} \left.-
\sum_{n = 1}^{ \infty}
{n\over 2}\left(\phi_n+\sqrt{2}{\sin(n\tau)  +\sin(n\rho)\over n
}\right)^2\right) \nonumber \\
&=&  e^{i ({4\phi_0-\tau-\rho)/2}}\exp \left(   i\sqrt{2}\sum_{n = 1}^{ \infty}
   \phi_n(e^{i n\tau}+e^{i n\rho}) -\sum_{n = 1}^{ \infty}{1\over n}
(\sin(n\tau)+\sin(n\rho))^2\right)\Psi_0 \, .
\nonumber \\
\eeqa
Now we note that
\begin{eqnarray}
\lefteqn{\sum_{n = 1}^{ \infty}{1\over n} (\sin(n\tau)+\sin(n\rho))^2}
\nonumber \\ 
 & = & \sum_{n = 1}^{ \infty}
   \left( {1-\cos(2n\tau)\over 2n}+{1-\cos(2n\rho)\over 2n}	+
	{2\sin(n\tau)\sin(n\rho)\over n}
\right)\\
& = &
   \sum_{n = 1}^{ \infty} {1\over n}+{1\over 4}\ln(2-2\cos 2\tau)+{1\over
4}\ln(2-2\cos 2\rho)
+{1\over 2} \ln\left({1-\cos(\tau+\rho)\over 1-\cos(\tau-\rho)}\right)\, ,
\nonumber
\end{eqnarray}
so that the shifted vacuum functional is
\beq
{1\over K^2} e^{i ({4\phi_0-\tau-\rho)/2}}\exp \left(
i\sqrt{2}\sum_{n = 1}^{ \infty}
   \phi_n(e^{i n\tau}+e^{i n\rho}) \right) \sqrt{ {1-\cos(\tau-\rho)\over
4|\sin\tau||\sin \rho | (1-\cos(\tau+\rho))} } \Psi_0\, ,
\eeq
where the $1/K^2$ comes from $\exp(-\sum {1\over n})=1/K^2$. This
factor of $1/K^2$
cancels the $K^2$ in $Q_f^2$, and the rest of the functional is
smooth inside the integral
in  (\ref{eq:qsq}), which now vanishes due to the antisymmetry of
$e^{i\pi \t(\tau-\rho)}$.

\section*{Acknowledgements}

Thanks to M.\ Douglas, I.\ Ellwood,
J.\ Harvey, G.\ Moore, L.\ Rastelli, A.\ Sen,
I.\ Singer, and B.\ Zwiebach for helpful discussions and
correspondence.  WT would like to thank the ITP, Santa Barbara, and
the ITP workshop on M-theory for support and hospitality during the
progress of this work.  The work of DJG was supported by the NSF under
the grants PHY 99-07949 and PHY 97-22022. The work of WT was supported
in part by the A.\ P.\ Sloan Foundation and in part by the DOE through
contract \#DE-FC02-94ER40818.

\normalsize

\bibliographystyle{plain}

\end{document}